\documentclass[fleqn,usenatbib]{mnras}

\usepackage{newtxtext,newtxmath}

\usepackage[T1]{fontenc}
\usepackage{ae,aecompl}

\usepackage{bm}
\usepackage{caption, subcaption}
\usepackage{graphicx}	
\usepackage{amsmath}	
\usepackage{amssymb}	
\usepackage{color}

\graphicspath{{./Plots/}}

\title[Contaminant-Free \textit{Gaia}-\textit{WISE} Galactic Plane Matches]{A Contaminant-Free Catalogue of \textit{Gaia} DR2-\textit{WISE} Galactic Plane Matches: Including the Effects of Crowding in the Cross-Matching of Photometric Catalogues}

\author[Tom J. Wilson and Tim Naylor]{
Tom J. Wilson,$^{1}$\thanks{E-mail: twilson@astro.ex.ac.uk}
and Tim Naylor$^{1}$
\\
$^{1}$School of Physics, University of Exeter, Stocker Road, Exeter EX4 4QL, UK\\
}

\date{Accepted XXX. Received YYY; in original form ZZZ}

\pubyear{2018}

\begin{document}
\label{firstpage}
\pagerange{\pageref{firstpage}--\pageref{lastpage}}
\maketitle

\begin{abstract}
Faint, hidden contaminants in the point-spread functions (PSFs) of stars cause shifts to their measured positions. Wilson \& Naylor (2017) showed failing to account for these shifts can lead to a drastic decrease in the number of returned catalogue matches in crowded fields. Here we highlight the effect these perturbations have on cross-matching, for matches between \textit{Gaia} DR2 and \textit{WISE} stars in a crowded Galactic plane region. Applying the uncertainties as quoted to Gaussian-based astrometric uncertainty functions (AUFs) can lead, in dense Galactic fields, to only matching 55\% of the counterparts. We describe the construction of empirical descriptions for AUFs, building on the cross-matching method of Wilson \& Naylor (2018), utilising the magnitudes of both catalogues to discriminate between true and false counterparts. We apply the improved cross-matching method to the Galactic plane $\lvert b \lvert\ \leq 10$. We provide the most likely counterpart matches and their respective probabilities. We also analyse several cases to verify the robustness of the results, highlighting some important caveats and considerations. Finally, we discuss the effect PSF resolution has by comparing the intra-catalogue nearest neighbour separation distributions of a sample of likely contaminated \textit{WISE} objects and their corresponding \textit{Spitzer} counterpart. We show that some \textit{WISE} contaminants are resolved in \textit{Spitzer}, with smaller intra-catalogue separations. We have highlighted the effect contaminant stars have on \textit{WISE}, but it is important for all photometric catalogues, playing an important role in the next generation of surveys, such as LSST.
\end{abstract}



\section{Introduction}
\label{sec:intro}
One of the most basic measurements available in astrophysics is the broadband photometric detection. Catalogues containing such detections, of differing wavelength coverage, resolution, dynamic range, etc., are often used together to maximise scientific potential. This merging process is the ``cross-matching'' of the catalogues, through which detections across several surveys corresponding to the same astrophysical source are identified and combined. However, the differences between the surveys introduce difficulties when constructing a merged dataset, which must be accounted for in order to not introduce systematic effects.  

The simplest catalogue match is done purely by sky separation, in a proximity-based nearest neighbour scheme. In this case the corresponding detection to a given star is assumed to be the closest star in the other catalogue, provided it is within a critical cutoff radius. These radii can vary considerably, from very tight matches (e.g., 1", \citealp{Dong:2011aa}; 3", \citealp{Cutri:2012aa}; 6", \citealp{Theissen:2016aa}) to larger radii (e.g., 16.5", \citealp{2015AJ....150..182K}; 1', \citealp{2013ApJ...779...61M}). However, this scheme has several limitations. Its primary issue is that it does not consider the possibility that the closest object is not the correct object. Additionally, despite the fact that there might be an object in the second catalogue within the critical radius, the source in question could have properties that would place its detection outside of the dynamic range of the second catalogue.

To overcome these limitations, probability-based catalogue matching methods have been developed (e.g., \citealp{2018MNRAS.473.5570W}; \citealp{Sutherland:1992aa}; \citealp{Naylor:2013aa}; \citealp{Budavari:2008aa}; \citealp{Rutledge:2000aa}; and references therein). These fold in information about the astrometric precision of astrophysical detections, allow for the rejection of all potential counterparts, and allow for the acceptance of an object that is not necessarily the closest in sky separation. With the more complex formalism the certainty to which the detections' positions are known can be included. This can lead to the possibility that an object with a larger absolute separation from another source can have a smaller normalised sky offset -- the ratio of its sky separation to the uncertainty in its position -- than one that is detected closer to the source. To achieve these improvements requires the creation of probability density functions (PDFs) that describe the likelihood of stars being related -- or not -- based on their respective astrometric precisions and sky separation. This improves upon the static cutoff radius of the nearest neighbour match by taking into account the relative astrometric precision of the catalogues.

These PDFs change based on the assumptions made about their form. The naive assumption is usually made that the astrometic uncertainty functions (AUFs) of each object are described by a two-dimensional Gaussian, as detailed by Quetelet (summarised by \citealp{Herschel:1857aa}). The AUF is the PDF that represents our belief as to the location of the object given its observed position. \citet{2018MNRAS.473.5570W} formally describe the probability of two objects being counterparts to one another as the convolution of the two stars' AUFs. They therefore use a match probability based on the functional form of the AUFs in question. They apply the method to catalogues for which the assumption that the AUF can be described by a Gaussian holds reasonably well.

There are cases when we need to be more flexible in our description of the AUFs used in the probability-based matching process. There is a small but consistent thread in the literature highlighting the effect that source confusion -- the inability to distinguish flux from one source from the flux of a second source -- has on the properties of those sources (e.g., \citealp{2001AJ....121.1207H}). For \textit{Wide-field Infrared Survey Explorer} (\textit{WISE}; \citealp{Wright:2010aa}), \citet{2017MNRAS.468.2517W} concluded that confusion caused by the crowding of faint contaminant stars (caused by effects of finite pixel size or point-spread function width) was a significant source of systematics in the AUFs. The undetected contaminating stars inside a bright object's point-spread function (PSF) lead to an AUF with a long, non-Gaussian tail. These perturbations act on length scales much greater than the typical perturbation due to non-zero proper motion, perhaps the most common additional cause of systematic perturbation. In turn, some separations between likely counterparts become much larger than previously assumed, even after accounting for smaller scale perturbations such as proper motion. Therefore, when considering a catalogue with significant crowding, like \textit{WISE}, we cannot ignore the effect of contaminants on the measured positions. If ignored, the non-Gaussian tails to the AUFs will reduce our astrometric likelihoods to a sufficient level to result in probability-based matches that return significantly fewer counterparts than a simple nearest neighbour-based match. 

This work is split into two main parts. We provide a catalogue of matches between \textit{Gaia} Data Release 2 (DR2; \citealp{2016A&A...595A...1G}; \citealp{Brown2018}) and \textit{WISE} for the Galactic plane ($\lvert b \lvert\ \leq 10$), utilising the catalogue matching procedure described by \citet{2018MNRAS.473.5570W}. The matching process combines a flexible formalism of the AUFs describing the detections in each photometric catalogue with the inclusion of the photometric information from both catalogues. This allows for the assignment or rejection of counterpart pairings on both astrometric and photometric probabilities, providing robust pairings with a low false match rate. We adapt this method to include the perturbation from faint contaminant stars as described by \citet{2017MNRAS.468.2517W}. These results are presented in Section \ref{sec:gplanematch}. Additionally, we more generally describe the procedure for the implementation of the effects of contaminant perturbation in the AUF in Section \ref{sec:empauf}. This method allows the reader to flexibly model the effect of fainter sources blended into the PSF of any photometric catalogue.

The layout of the paper is as follows. Section \ref{sec:gaussauf} describes the probability-based matching of \textit{Gaia} and \textit{WISE} in the case where contamination is not taken into account. Section \ref{sec:empauf} details how to correct the AUF of a probability-based matching method empirically to include the effects of crowding, and applies the method to the \textit{Gaia}-\textit{WISE} matching case. We then detail the application of the probability-based matching method to \textit{Gaia} and \textit{WISE} for a large section of the Galactic plane in Section \ref{sec:gplanematch}, providing the most likely match and its corresponding probability, both of the match and of source contamination. This includes analysis of test cases, comparison with previous results, and a discussion of the photometric effects of crowding. Here we show that the additional matches that are astrometrically perturbed enough to be missed by a Gaussian probability-based match are flux contaminated by an average of 30\%. We also compare the \textit{WISE} matches to \textit{Spitzer} \citep{2004ApJS..154....1W}, showing that the higher angular resolution of \textit{Spitzer} sometimes allows for the resolving of the hidden \textit{WISE} contaminants. Section \ref{sec:discussion} provides a brief discussion of several implications these results have, highlights a few minor caveats to the data product, and discusses extensions to the methodology. Concluding remarks are then given in Section \ref{sec:conclusions}. Table \ref{tab:symbols} defines symbol usage in the paper.

\begin{table}
\begin{tabular}{c | l}
\hline
Symbol & Definition \\
\hline
$B$ & Magnitude density of sources at given magnitude\\
$c(m_\gamma, m_\phi)$ & PDF of counterpart magnitude relationship\\
$D$ & Differential source counts\\
$f_\phi(m)$ & PDF of unmatched catalogue $\phi$ stars\\
$F_\mathrm{contam}$ & Average flux contamination of sources\\
$G$ & PDF of two stars being related given an offset\\
$h_\phi$ & Astrometric uncertainty function of catalogue $\phi$\\
$l$, $b$ & Galactic sky coordinates\\
$m$ & The magnitude of a given star\\
$m_i$ & Magnitude of differential source count break\\
$N_\mathrm{c}$ & Counterpart number density\\
$N_\phi$ & Number density of unmatched stars in catalogue $\phi$\\
$N$, $N_i$ & Geometric number density normalisation constants\\
$P_\mathrm{match}$ & Counterpart match probability\\
$P_\mathrm{contam}$ & Probability of source being contaminated\\
$r$ & Radial separation\\
$R$ & PSF radius\\
$\mathcal{R}_Y$ & Radius defining circular PDF integral\\
$T$ & Number of stars in a given magnitude range\\
$U$ & Number of objects in circle of given radius\\
$W$ & Average number of PSF contaminants\\
$x$, $y$ & Cartesian sky offsets\\
$Y$ & Fraction of PDF integral\\
$z$, $z_i$ & Geometric scaling laws\\
$\alpha$, $\delta$ & Celestial coordinates\\
$\gamma$ & A Catalogue\\
$\Delta m$ & Given magnitude offset from central source\\
$\Delta m_\mathrm{max}$ & Maximum magnitude offset\\
$\eta$ & Photometric likelihood ratio\\
$\theta$ & Position angle of sky axes\\
$\mu$ & Astrometric proper motion\\
$\xi$ & Astrometric likelihood ratio\\
$\rho$ & Correlation of celestial sky axis uncertainties\\
$\sigma_\alpha$, $\sigma_\delta$ & Celestial sky axis uncertainties\\
$\sigma_\mathrm{pure}$ & Intrinsic astrometric centroiding uncertainty\\
$\phi$ & A catalogue\\
$\psi$ & A contamination hypothesis\\
$\omega$ & A contamination hypothesis\\
\hline
\end{tabular}
\caption{Table showing the definition of symbols used throughout this paper.}
\label{tab:symbols}
\end{table}

\begin{table*}
\centering
\begin{tabular}{c | c | c}
\hline
Catalogue & Flag & Criteria\\
\hline
\textit{Gaia} DR1 & Poor Fit & astrometric\_excess\_noise $>$ 2.375mas and astrometric\_excess\_noise\_sig $>$ 2\\
 & Low Quality & astrometric\_excess\_noise $>$ 2.375mas and astrometric\_excess\_noise\_sig $\leq$ 2; or\\
 & & astrometric\_n\_good\_obs\_al + astrometric\_n\_good\_obs\_ac $<$ 60; or matched\_observations $\leq$ 8\\
 \hline
\textit{Gaia} DR2 & Poor Fit & sqrt(astrometric\_chi2\_al/(astrometric\_n\_good\_obs\_al$\,$-$\,$5)) $>$ 1.2$\,$max(1, $\exp\left(-0.2(\mathrm{phot\_g\_mean\_mag}\,-\,19.5)\right)$)$^1$\\
 & Low Quality & astrometric\_excess\_noise $>$ 1mas$^1$ and astrometric\_excess\_noise\_sig $\leq$ 2; or\\
 & & astrometric\_n\_good\_obs\_al $<$ 60; or astrometric\_matched\_observations $\leq$ 8\\
 \hline
\textit{WISE} & Non-stellar & ``Contam'' flag is either ``D'', ``P'', ``H'', or ``O''; or ``ext'' flag is 2, 3, 4, or 5\\
 & Outside Dynamic Range & ``Phqual'' flag is ``X'' or ``Z''; or ``detbit'' == 0; or Mag == NaN; or ``sat'' flag $>$ 0; or $\sigma_{\mathrm{Mag}}$ == NaN\\
 & Low Quality & ``Phqual'' flag is ``U''; or ``Contam'' flag is ``d'', ``p'', ``h'', or ``o''; or ``ext'' flag is 1; or\\
 & & ``var'' flag is $>5$ or ``n''; ``nblend'' flag is $>$ 3; or ``moonlev'' $>$``1''\\
\hline
\textit{Spitzer} & Outside Dynamic Range & Saturation Flag is set, or Artefact of Wing Saturation flag is set\\
 & Low Quality & Dark Current flag is set, Flat Field flag is set, Latent Image flag is set, Bad Pixel flag is set,\\
 & & In-band or Cross-band Merge Confusion flags are set, or Edge of Frame flag is set\\
 \hline
\end{tabular}
\caption{Table showing the various flags for non-stellarity, artefacts, detection and photometric quality for the catalogues used. $^1$See appendix C of \citet{Lindegren2018} for more details.}
\label{tab:flags}
\end{table*}

\begin{figure*}
    \centering
    \includegraphics[width=\textwidth]{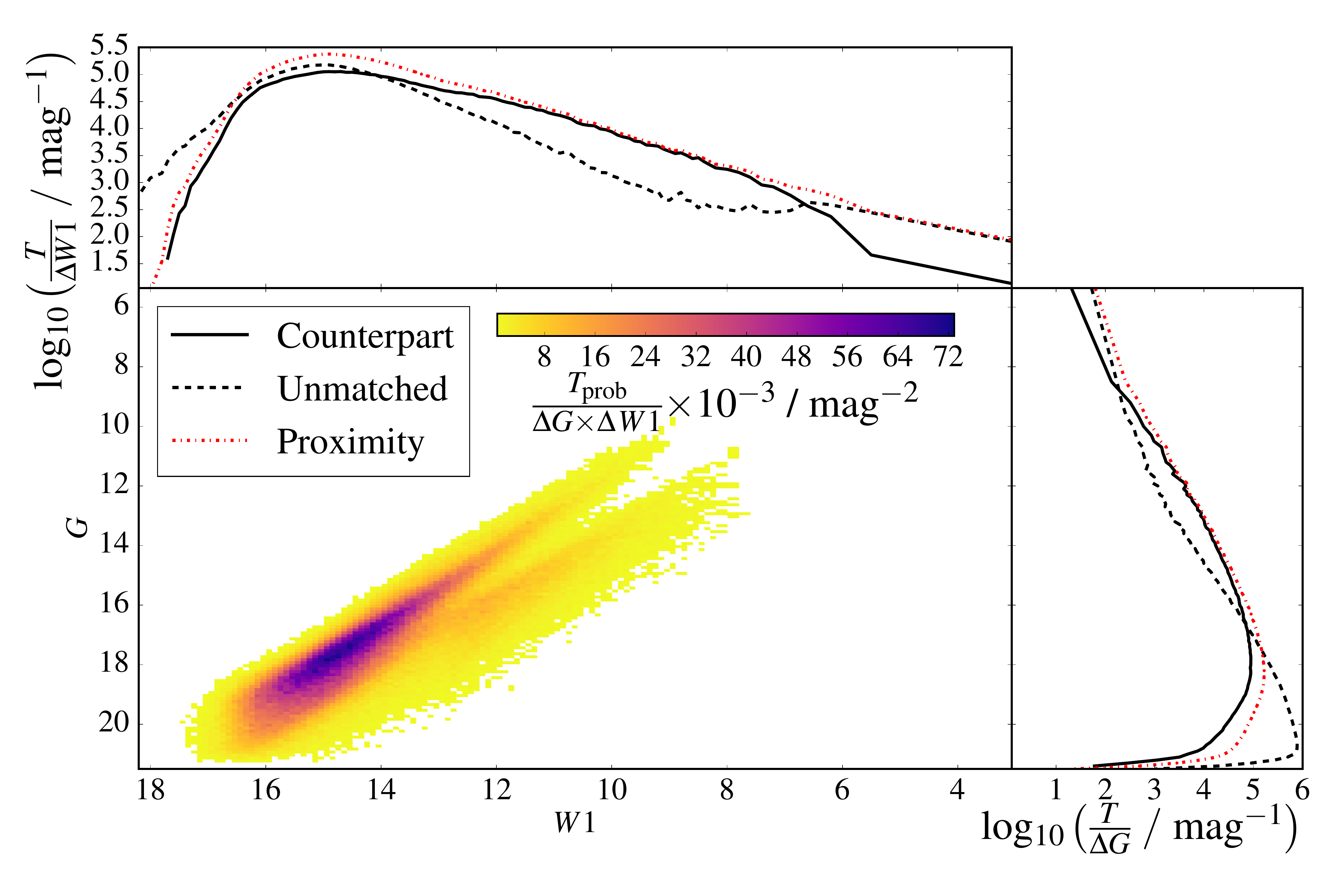}
    \caption{The number density of matched objects between \textit{Gaia} and \textit{WISE} for a 42 square degree region of the Galactic plane. The main panel shows a 2D histogram of the number density of objects -- number of sources, $T$, per unit \textit{WISE} magnitude per unit \textit{Gaia} magnitude -- for objects with both $G$ and $W1$ detections as a function of $G$ and $W1$. The two inset panels show the number density of objects as functions of $G$ or $W1$ magnitude alone. The inset panels show the results of 3" nearest neighbour matches in a red dash-dotted line, probability-based counterparts in a solid black line, and probability-based unmatched ``field'' stars in a black dashed line. Using a Gaussian to represent the AUF results in matches for 56\% of the nearest neighbour matches. Only bins with densities $\geq500\mathrm{mag}^{-2}$ are displayed in the main panel.}
    \label{fig:gwgauss}
\end{figure*}

\section{The Gaussian Astrometric Uncertainty Function}
\label{sec:gaussauf}
Before we can quantify the significance of the inclusion of perturbations in the description of the AUFs, we must first discuss the matches obtained without their consideration. Therefore our first choice of AUF should be the most obvious, the assumption made most often in probability-based matching: that the probability of two detections of a source being at a given separation is entirely described by a Gaussian. In this section we will describe the results of matching \textit{WISE} to \textit{Gaia} in a crowded region of the Galactic plane under the assumption of a purely Gaussian AUF.

\subsection{Constructing the Gaussian AUF}
\label{sec:constructgaussauf}
When using a probability-based matching method, the astrometric PDF is usually assumed to be a two-dimensional zero-centered Gaussian with covariance matrix

\begin{equation}
\bm{\Sigma} = \left( \begin{array}{cc} \sigma_\alpha^2 & \rho\sigma_\alpha\sigma_\delta \\ \rho\sigma_\alpha\sigma_\delta & \sigma_\delta^2 \end{array} \right),
\label{eq:covariancematrix}
\end{equation}
where $\sigma_\alpha$ and $\sigma_\delta$ are the convolved \textit{Gaia}-\textit{WISE} uncertainties in the two orthogonal sky directions (Right Ascension and Declination, respectively) and $\rho$ is the correlation between the two. $G$ is then

\begin{align}
\begin{split}
G(\Delta \alpha, \Delta \delta) &= \frac{\exp{\left(-\frac{1}{2\sqrt{1 - \rho^2}}\left(\frac{(\Delta \alpha)^2}{\sigma_\alpha^2} + \frac{(\Delta \delta)^2}{\sigma_\delta^2} - \frac{2\rho \Delta \alpha\Delta \delta}{\sigma_\alpha\sigma_\delta}\right)\right)}}{2\pi\sigma_\alpha\sigma_\delta\sqrt{1-\rho^2}},
\label{eq:bivariatepdfs}
\end{split}
\end{align}
where $\Delta \alpha$ and $\Delta \delta$ are the orthogonal sky axis offsets between the respective \textit{Gaia} objects and \textit{WISE} sources, including the cosine of the declination which converts our right ascension separations entirely to seconds of arc. 

\subsection{The Effects of the Gaussian AUF on \textit{Gaia}-\textit{WISE} Matches}
\label{sec:gaussaufeffect}
To test the effect the AUF has on the resulting pairings, we matched \textit{Gaia} DR2 stars against \textit{WISE} stars. For a 42 square degree region of the Galactic plane, $131 \leq l \leq 138$, $-3 \leq b \leq 3$, we filtered the catalogues for poor quality, non-stellarity and non-detections as described in Table \ref{tab:flags}, using the minimum recommended DR2 filtering cuts, as described in appendix C of \citet{Lindegren2018}. We also applied a proper motion correction to the \textit{Gaia} dataset, as given by equation \ref{eq:pmdrift}, accounting for the epoch difference between the two datasets where proper motions were available. We used the probability-based matching process of \citet{2018MNRAS.473.5570W}. In all cases it is assumed that $G$ (the convolution of the AUF of each source), and any defining merging/cutout radii $\mathcal{R}_Y$ (the circle radius inside which the integral of $G$ is equal to $Y$), are Gaussian. These functions are described in further detail by \citet{2018MNRAS.473.5570W}. We also use this assumption when using the photometric information available in the catalogues to construct $c$ and $f$ and evaluate our photometric probabilities, also detailed by \citet{2018MNRAS.473.5570W}. 

The results of this cross-match are shown in Figure \ref{fig:gwgauss}, for matches with a probability $P \geq 0.5$ (see \citealp{2018MNRAS.473.5570W} for details on how the match probabilities are calculated). The counterparts the cross-matching process returns have magnitudes which lie in a sensible region of the $G-W1$ magnitude-magnitude plane (main panel, Figure \ref{fig:gwgauss}). As expected, the density of matches increases towards fainter $G$ and $W1$ magnitudes, with \textit{Gaia} magnitudes typically 1-4 magnitudes fainter than the \textit{WISE} passbands. We also recover the dwarf-giant separation towards brighter magnitudes ($W1 \leq 12$).

However, as shown by the side panels of Figure \ref{fig:gwgauss}, the assumption that the positional uncertainties are described by a Gaussian results in only 56\% the matches that were returned using a 3" nearest neighbour-based matching procedure. Assuming a \textit{Gaia} stellar density of $2\times10^4\, \mathrm{deg}^{-2}$ in the area of the Galactic plane in question \citep{2017MNRAS.468.2517W} we get a false match rate on the order of 4\%. We therefore cannot explain our additional nearest neighbour-based matches entirely as false matches, as we expect the removal of approximately 4\% of our nearest neighbour matches but reject 44\% of these matches in our probability-based match. This order-of-magnitude increase in rejection rate must have a different explanation.

\begin{figure}
    \centering
    \includegraphics[width=\columnwidth]{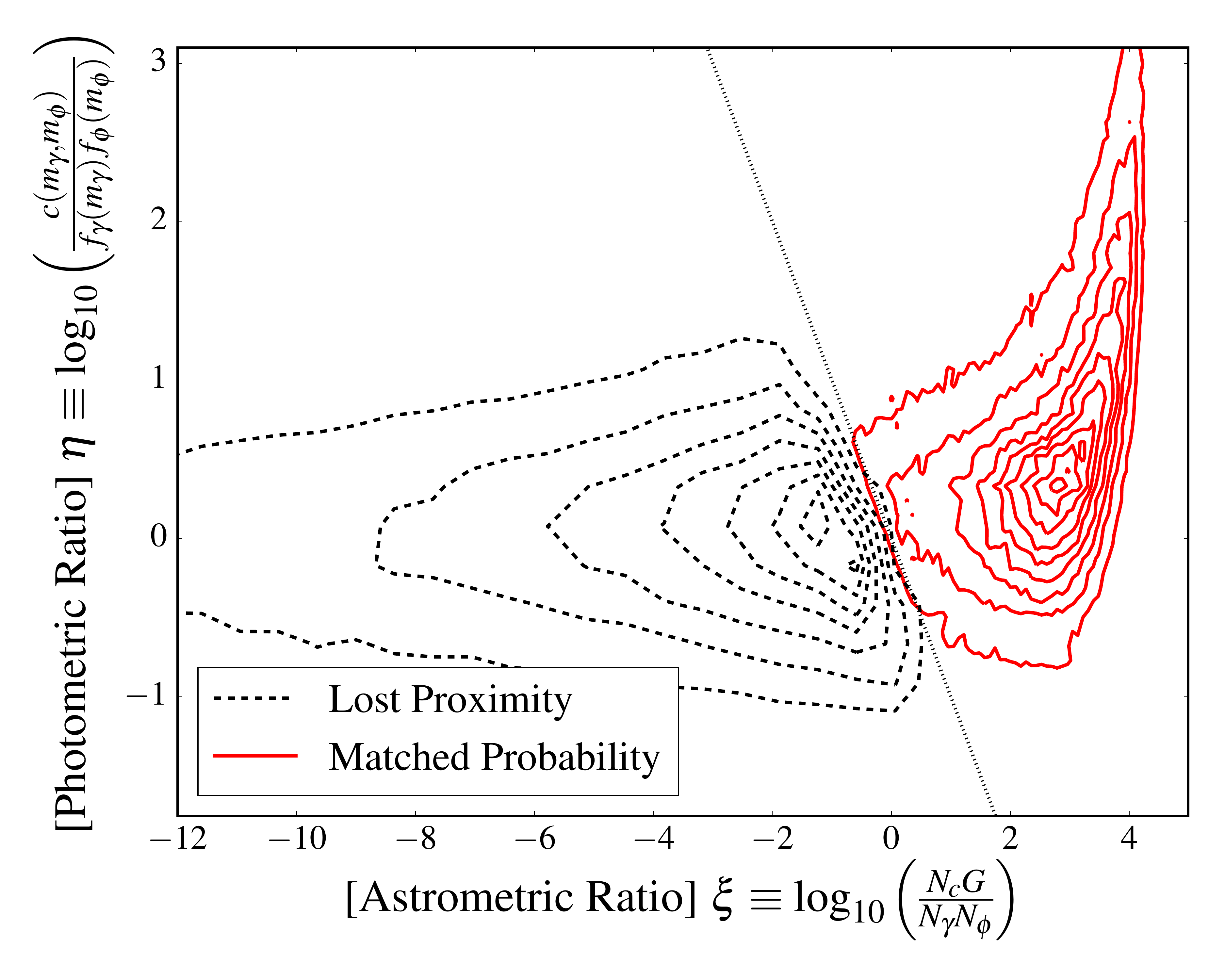}
    \caption{The photometric and astrometric likelihood ratios of \textit{Gaia} matches for a 42 square degree region of the Galactic plane, under the assumption that the distribution of separations follows a Gaussian distribution. The density of objects returned as pairs by the probability-based match is shown as red solid contours. Objects paired by a 3" nearest neighbour match but returned as unrelated in the probability-based match are shown as the dashed black contours. The lost matches are below the equal probability dotted line, $\xi + \eta = 0$, but at a lower astrometric likelihood ratio, with a roughly constant photometric likelihood ratio $\eta\simeq0.3$. This implies the pairings are more likely than not based on photometry arguments, but are lost due to the assumptions made about $G$.}
    \label{fig:likeratiogauss}
\end{figure}
\vspace*{3\baselineskip}
Considering the likelihood ratios (e.g., \citealp{Sutherland:1992aa}) of the astrometric and photometric halves of the equations used in the probability-based matches (see \citealp{2018MNRAS.473.5570W} for more details) shows the reason for the loss of these nearest neighbour matches. The photometric likelihood ratio, $\eta$, is defined as the logarithm of the ratio of the counterpart probability density, $c$, to the likelihood of the two unmatched densities, $f_\gamma\cdot f_\phi$. Equivalently, the astrometric likelihood ratio, $\xi$, logarithmically balances the astrometric counterpart probability density, $N_\mathrm{c}G$, with the probability density of two unrelated objects, $N_\gamma\cdot N_\phi$.

As shown in Figure \ref{fig:likeratiogauss}, the majority of the objects matched return both a high astrometric ($\xi\geq0$) and photometric ($\eta\geq0$) likelihood ratio; they are more likely than not to be matched on both spatial and magnitude grounds. At very high matched object astrometric likelihood ratios ($\xi\geq2$) the photometric likelihood is very high as well. At lower (albeit still more than equal probability) astrometric likelihood ratios the photometric likelihood ratio of these matched objects plateaus at $\eta\simeq0.3$.

Also shown in Figure \ref{fig:likeratiogauss} are the likelihood ratios for any pairs that are nearest neighbour matched within 3" but not returned as a pair by the probability-based match. These are all below the equal likelihood ratio line, defined as being $\xi + \eta = 0$. However, they still follow $\eta\simeq0.3$, implying a photometric likelihood ratio higher than equal chance, and no lower on average than the returned probabilistic matches. Therefore, the matches are failing to be returned due to their astrometric likelihood ratio, which rapidly decreases to several orders of magnitude below equal likelihood.

\section{The Empirical Astrometric Uncertainty Function}
\label{sec:empauf}
Since we are losing our probability-based matches on purely astrometric arguments, we must reconsider our definition of $G$ to correct for the missing $\simeq 45\%$ of the nearest neighbour counterparts. To achieve this, we can construct empirical AUFs, based on the distribution of separations for a given area of the sky. This allows us to account for varying levels of crowding seen at varying longitudes and latitudes throughout the sky.

\subsection{Constructing the Empirical AUF}
\label{sec:constructempauf}
To model the AUF of perturbed stars, some simple numerical models are required, as discussed by \citet{2017MNRAS.468.2517W}. In these we simulate the effects of a given stellar density, recording the positions of stars inside the bright, central star's PSF, including the effects of stars well below the completeness limit of our survey. This distribution is then combined with the intrinsic positional uncertainty of the given star. The resulting PDF, for the offset of the star from its true position, is the perturbed-star AUF we require.

First we must obtain the distribution of physical perturbations of a star in the stellar field in question. For this example we assume, following \citet{2017MNRAS.468.2517W}, that the stellar density of objects as a function of magnitude in a given filter, $D$, follows a geometric series. This density gives the number of stars per unit magnitude per unit area, $D = N z^m$, at magnitude $m$. Here $N$ is the stellar density (per unit magnitude per unit area) of the field at zeroth magnitude, and $z$ is the geometric scaling factor which dictates the rate of increase of the stellar density with decreasing brightness. In addition, since we are only interested in the stars within our PSF circle, we must include the term for the circle area, giving us an effective ``magnitude density'' at the magnitude of our star,

\begin{equation}
B = N z^m \pi R^2,
\label{eq:Bdensity}
\end{equation}
where $R$ is the radius of the PSF circle. For this radius we use the Rayleigh criterion \citep{1880MNRAS..40..254R} of the telescope

\begin{figure}
    \centering
    \includegraphics[width=\columnwidth]{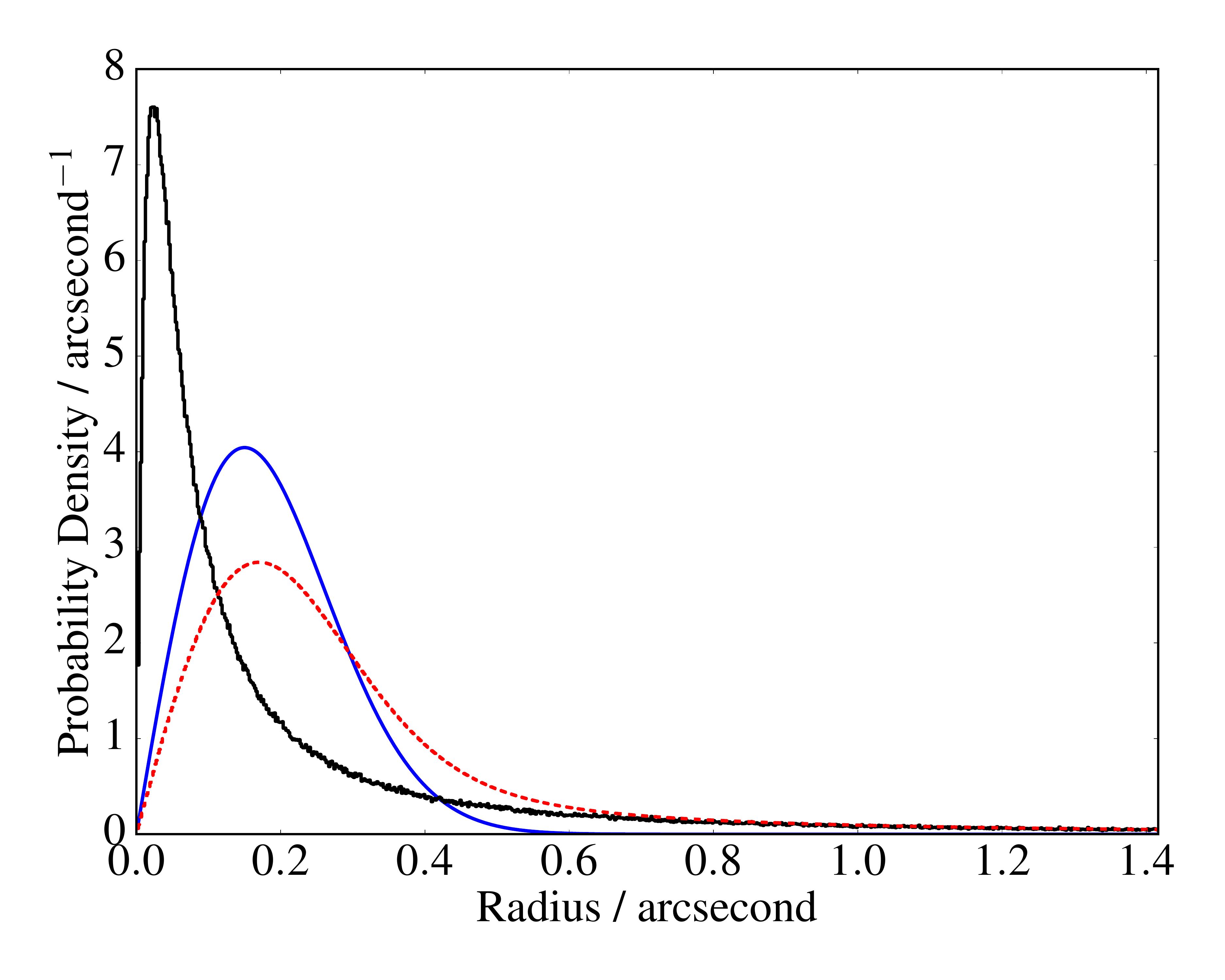}
    \caption{An example numerical AUF, for $N = 0.5\,\mathrm{mag}^{-1}\,\mathrm{deg}^{-2}, z = 2, \sigma_\alpha = 0.15", m = 12$. In this case the contaminating stars are drawn from a distribution from equal brightness down to a magnitude difference $\Delta m_\mathrm{max} = 10$, or a flux ratio of 0.0001. The black histogram shows the distribution of central star perturbations from the origin, caused by the flux-weighted average positions of the contaminating stars. The blue solid line shows the ``pure'' Gaussian from which the measured position would be naively drawn, represented here by a Rayleigh distribution, the transformation of a two-dimensional Gaussian to one-dimensional radial coordinates. The red dashed line shows the convolution of the two, giving the resulting AUF.}
    \label{fig:numericaufexample}
\end{figure}

\begin{equation}
R = 1.185 \times \mathrm{FWHM}
\label{eq:airyradius}
\end{equation}
as described by \citet{1835TCaPS...5..283A}, where FWHM is the full width at half maximum of the telescope PSF. To build up a sample PSF contamination we must evaluate the chance of a contaminant star of given magnitude offset (i.e., with a certain flux ratio) $\Delta m$ relative to our bright central source (of magnitude $m$) being in our PSF circle. The average number of stars of each faint magnitude slice $m + \Delta m$ is given by integrating the magnitude density across the bin width (d$m$),

\begin{align}
\begin{split}
P_B(m + \Delta m) &= \!\!\!\!\!\!\!\!\!\int\limits_{m+\Delta m}^{m+\Delta m + \mathrm{d}m}\!\!\!\!\!\!\!\!\!\! B(m{'})\, \mathrm{d}m{'} = \!\!\!\!\!\!\!\!\!\!\int\limits_{m+\Delta m}^{m+\Delta m + \mathrm{d}m}\!\!\!\!\!\!\!\!\!\!N z^{m{'}} \pi R^2\, \mathrm{d}m{'}\\ 
&= \frac{N z^{m+\Delta m} \left(z^{\mathrm{d}m} - 1\right)}{\log(z)} \times \pi R^2.
\label{eq:stellardensity}
\end{split}
\end{align}
Using this typical source count, we draw from a Poission distribution the expected number of stars in the PSF at this magnitude slice. If non-zero, these stars are randomly distributed in $\theta$ and $r^2$ space (to account for the additional $r$ term in the unit circle area term). These are then converted to Cartesian coordinates. This is repeated for $\Delta m = 0$ to $\Delta m = 10$ in steps of $\mathrm{d}m = 0.025$. Once all magnitude slices have had stars randomly drawn and distributed, the flux-weighted average $x$ and $y$ positions are recorded, and converted back to a radius as $r = \sqrt{x^2 + y^2}$. 

This sampling of contaminant star brightnesses and radial offsets is repeated for a million unique test PSFs, each time stepping through $\Delta m$. This results in a distribution of offsets, which is then converted to a PDF. This perturbation PDF, $h_\mathrm{offsets}$, is then convolved with the Gaussian, $h_\mathrm{pure}$, of the intrinsic positional uncertainty of the central star, $\sigma_\mathrm{pure}$, to produce a numerical AUF for a given stellar density, brightness, and positional uncertainty. Mathematically, this is given by

\begin{equation}
h_\mathrm{tot} = h_\mathrm{pure} * h_\mathrm{offsets}.
\label{eq:mathsempconv}
\end{equation}
An example of such an AUF is shown in Figure \ref{fig:numericaufexample}. The intrinsic Gaussian AUF (blue solid line) is convolved with the distribution of perturbations (black histogram), resulting in an empirical AUF (red dashed line) that includes the effects of the blending of faint contaminant stars into the PSF of the central source on its astrometric position.

In cases of very low crowding, either through high angular resolution and thus small $R$, low source densities and thus low $N$, or through bright central magnitude and thus low $z^m$, the central offset will tend to zero in most numerical simulations. In these cases $h_\mathrm{offsets}$ reduces, effectively, to a delta function ($\delta_\mathrm{offsets}$). For these low crowding cases the AUF is simply the intrinsic Gaussian AUF in the absence of any contamination,

\begin{equation}
h_\mathrm{tot} = h_\mathrm{pure} * \delta_\mathrm{offsets} = h_\mathrm{pure}.
\label{eq:mathsempconv_delta}
\end{equation}
We therefore simply have to convolve our ``offset'' AUF component with our ``pure'' AUF component, regardless of the levels of contamination suffered by any individual source.

\subsection{The Dependences of Empirical AUF Construction}
\label{sec:empaufoveralldependence}
Our parameterisation of the level to which contaminant stars affect our astrometric position, $B$ (see equation \ref{eq:Bdensity}), is dependent on three further parameters: first, the brightness of the central source, $m$; second, the overall source density in the region of sky in question, $N$; and third, a description of the increase in source counts with increasing magnitude, $z$. We must therefore explore the effects of this parameterisation before we can construct our empirical AUFs across a large area of the Galactic plane.

\subsubsection{The Dependence of $N$ and $z$ on $l$ and $b$}
\label{sec:empauflbdependence}
First we must decide whether $z$ should be described as a function of $l$ and $b$. Dense regions of gas (e.g., molecular clouds) will, in theory, cause differential extinction preferentially extincting more distant and fainter stars. However, if we could assume a constant geometric scaling (e.g., \citealp{2010ApJ...724..182C}), it would greatly simplify the creation of empirical AUFs, allowing us to simply scale our total source density through our choice of $N$. 

To test this, we initially fitted the differential source counts in a small region of Galactic plane, $134 \leq l \leq 134.2$, $2 \leq b \leq 2.2$, small enough that there should be limited effects from differing source densities across star forming regions, using $Nz^m$. We found $z = 1.978$. As this value is very close to 2, we tentatively adopt $z = 2$ as our canonical geometric scaling law, but first must ensure that this value is appropriate across a variety of differential crowdings. 

\begin{figure}
    \centering
    \includegraphics[width=\columnwidth]{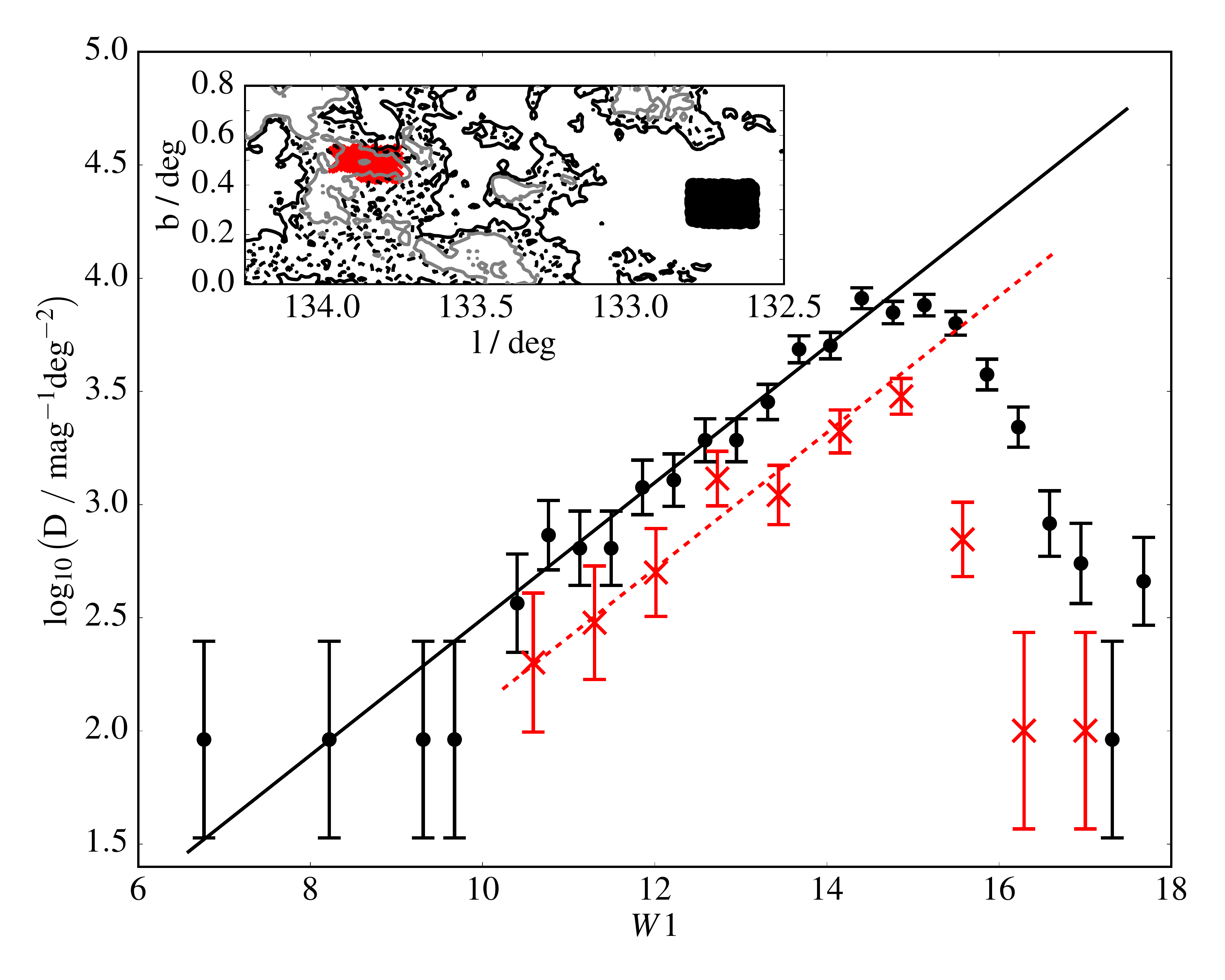}
    \caption{The effect of differing stellar densities on source counts. Inset panel shows the spatial distribution of two small patches of \textit{WISE} stars. Contours denote levels of integrated $^{12}$CO brightness temperature, 7.23/13.45/19.87 K$\,$km$\,$s$^{-1}$ respectively, used here as a proxy for column density. Black dots show stars in a region of low integrated brightness temperature, while red crosses mark out a separate region, this time in a higher column density. Main panel shows differential source counts for the two regions. Red dashed and solid black lines represent the best fits to the two datasets respectively, $D = Nz^m$, assuming a geometric scaling law $z=2$. The two fits have values of $N = 0.127\,\mathrm{mag}^{-1}\,\mathrm{deg}^{-2}$ and $N = 0.304\,\mathrm{mag}^{-1}\,\mathrm{deg}^{-2}$, respectively.}
    \label{fig:zlbhist}
\end{figure}
 
The inset panel of Figure \ref{fig:zlbhist} shows a small region of the Galactic plane, $132.5 \leq l \leq 134.25$, $0 \leq b \leq 0.8$. We have selected two smaller regions of interest, with different source counts and column densities, represented here through the proxy of $^{12}$CO integrated brightness temperature, using the FCRAO OGS survey \citep{1998ApJS..115..241H}. Shown in black are stars in a region with low column density, whereas the red data are stars in the line of sight of a molecular cloud, affecting the differential source counts. 

The differential source counts for the two regions are shown in the main panel of Figure \ref{fig:zlbhist}, with the best fits to the data, assuming a scaling law of $z = 2$. In both cases the scaling law fits well, with a simple reduction in $N$ for the region of higher column density. We find this relationship fits well across multiple photometric catalogues with differing spatial resolution and wavelength coverage, and therefore suggest $z = 2$ as the invariant bright geometric scaling law across all catalogues and sky positions (see Section \ref{sec:empaufzmdependence} for further discussion). However, the intrinsic source density does vary with sky position, and $N$ is still parameterised by the local sky density.

\subsubsection{The Dependence of Differential Source Counts on Central Star Brightness}
\label{sec:empaufzmdependence}
The assumption was made in Section \ref{sec:constructempauf} that differential source counts as parameterised by \citet{2017MNRAS.468.2517W} can be extrapolated below the completeness limit of a given survey. However, as shown by, e.g., \citet{1980ApJ...238L..17B}, there is a decrease in the count rate of the very faintest objects. This effect has several sources; one of the primary causes of the decrease in sources at these magnitudes is the edge of the Galaxy, beyond which stellar densities are much diminished. This turnover means that our previous extrapolation of count rates for a central star of e.g., $W1 = 17$ down a further 10 magnitudes would lead to unphysical contamination fractions; this effect is discussed further in Section \ref{sec:invispertub}. We must therefore re-parameterise our differential source count model to account for this issue at faint magnitudes.

\begin{figure}
    \centering
    \includegraphics[width=\columnwidth]{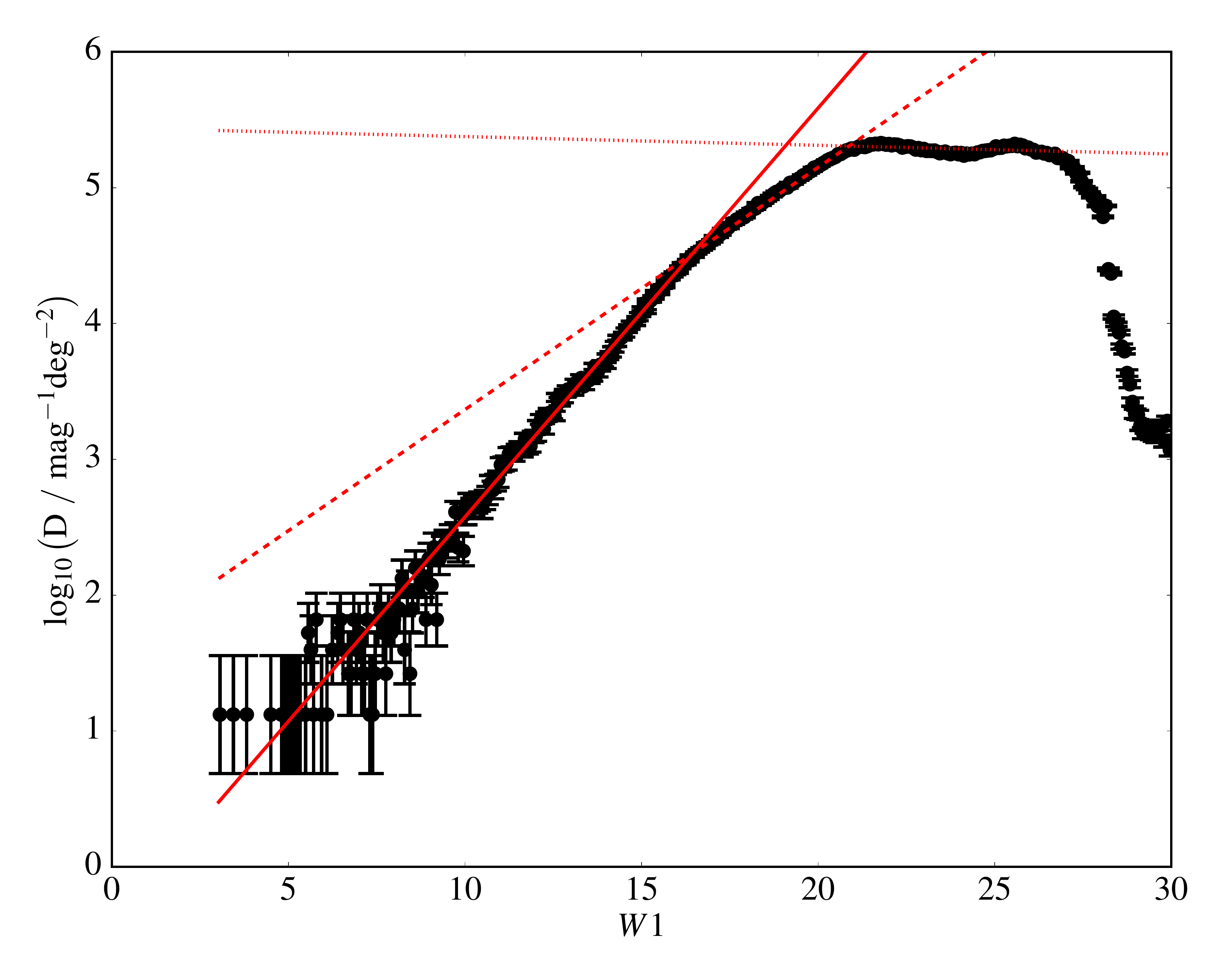}
    \caption{TRILEGAL differential source counts for one square degree centered on $l = 133, b = 0$. Also shown are three fits, $D = N z^m$, to three different parts of the source counts. The solid red line shows the fit to sources $W1 \leq 16.5$, fixed at $z_1=2$, for which the best fit value is $N = 0.367\,\mathrm{mag}^{-1}\,\mathrm{deg}^{-2}$. The dashed red line shows the fit to $16.5 \leq W1 \leq 21$, fixed such that the crossover differential source counts are consistent at $W1 = 16.5$, resulting in a $z_2 = 1.51$. The dotted red line shows the fit to $21 \leq W1 \leq 26$, again fixed such that the crossover differential source counts agree at $W1 = 21$, resulting in $z_3 = 0.99$.}
    \label{fig:trilegalcounts}
\end{figure}

To analyse the differential source counts below the \textit{WISE} completeness limit, a TRILEGAL\footnote{http://stev.oapd.inaf.it/cgi-bin/trilegal} \citep{Girardi2005} simulation for one square degree of the Galactic plane centered on $l = 133,\ b = 0$ was obtained. The $W1$ differential source count for the region is shown as black error bars in Figure \ref{fig:trilegalcounts}. Also shown in Figure \ref{fig:trilegalcounts} are three red lines, representing a geometric scaling law parameterisation of the source counts. This multiple law parameterisation is defined by a number of scaling laws (in this case $z_1 = 2$, $z_2 = 1.51$, $z_3 = 0.99$) and crossover magnitudes ($m_2 = 16.5$, $m_3 = 21$). We define each subsequent scaling law normalisation beyond the first as being $N_{i+1} = N_i z_i^{m_{i+1}} / z_{i+1}^{m_{i+1}}$, in which the effective differential source counts for each parameterisation is the same at the crossover magnitude. The entire parameterisation therefore depends solely on the initial normalisation density $N_1 \equiv N$. This is still the source density defined by stars at the bright end of the catalogue, typically easily obtained from detected source counts above the survey completeness limit.

While we have shown that a multi-scaling law parameterisation can remain a reasonable approximation to the differential source counts of a catalogue, we choose to no longer describe our differential source counts analytically. For the remainder of this paper we choose instead to build our synthetic Galactic \textit{Gaia}-\textit{WISE} match offset distributions using TRILEGAL simulations to parameterise $z(m)$. We therefore update the method described in Section \ref{sec:constructempauf} to utilise the simulated TRILEGAL stellar population. $D$ is created as a histogram of the simulated magnitude distribution (cf. Figure \ref{fig:trilegalcounts}), and then the expected number of contaminant stars at each magnitude step is calculated as

\begin{equation}
P_B(m + \Delta m) = \frac{N_\mathrm{empirical}}{N_\mathrm{TRILEGAL}}D(m + \Delta m) \times \pi R^2 \times \mathrm{d}m,
\label{eq:pbfromtrilegal}
\end{equation}
where $N_\mathrm{empirical}$ is the local bright-magnitude normalising density of the catalogue in question, and $N_\mathrm{TRILEGAL}$ is the equivalent normalising density of the simulated data. This ratio is a simple correction factor to re-normalise the relative counts to those of the data; the important information the simulated magnitude differential source counts provide is $z(m)$.

This method is used to construct the AUFs used to evaluate the \textit{Gaia}-\textit{WISE} matches in Section \ref{sec:gplanematch}. However, it should be noted that there may be certain cases where such simulations may not be available or relevant, in which case the power law parameterisation may be the preferred choice. This is discussed further in Section \ref{sec:aufextensions} for the case of faint sources out of the plane of the Galaxy, where extragalactic sources dominate the differential source count.

\subsection{Applying a Empirical AUF to \textit{Gaia}-\textit{WISE} Separations}
\label{sec:empauftest}
Now that we have a complete description of the differential source counts in a given filter, including effects below the catalogues' sensitivity, we can construct new AUFs. Each empirical AUF is uniquely described by three parameters: $N$, the geometric scaling normalisation of the bright part of the scaling law; $m$, the magnitude of the central source; and $\sigma_\mathrm{pure}$, the intrinsic uncertainty of the centroiding of the central source in the absence of crowding. We calculate $N$ by obtaining the number of \textit{WISE} objects in the range $9 \leq W1 \leq 14$ within 15 arcminutes of each \textit{WISE} source, $U$, and solving the equality

\begin{equation}
U = \int\limits_{m{'}=9}^{14}\!N z^{m{'}} \mathrm{d}m{'} \times \pi \times (15')^2,
\label{eq:solveforn}
\end{equation}
assuming $z=2$. Once we have $N$ and $m$, we can increment through each contaminant star magnitude, calculating $P_B$ (cf. equation \ref{eq:pbfromtrilegal}, or equation \ref{eq:stellardensity} for the bright scaling law limiting case) and drawing contaminant stars to place within the PSF. The flux-weighted average of all of the stars in a given PSF can be found, and the process repeated, as described in Section \ref{sec:constructempauf}.

\begin{figure}
    \centering
    \includegraphics[width=\columnwidth]{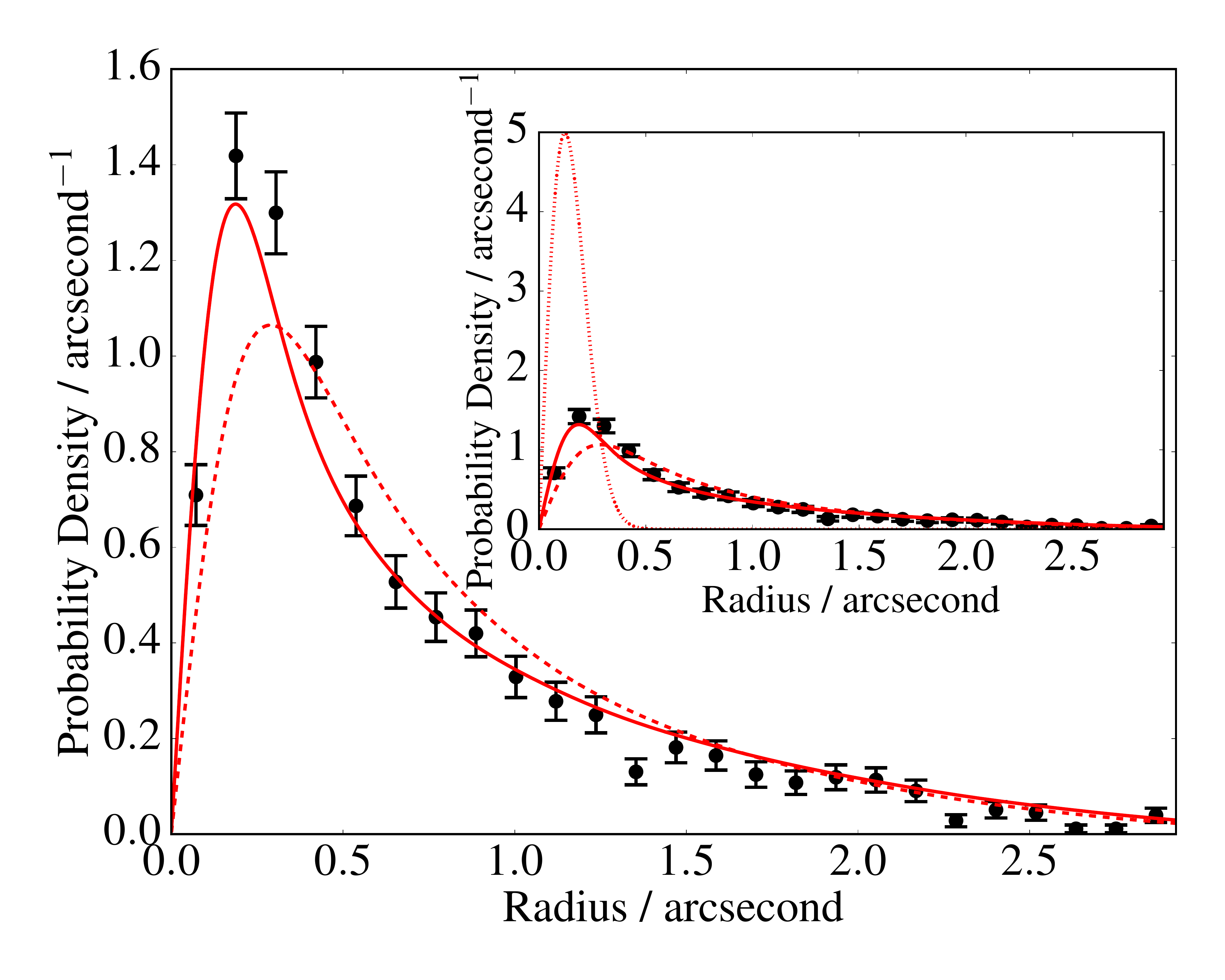}
    \caption{The distribution of separations between \textit{Gaia} and \textit{WISE} objects for 42 square degree region of the Galactic plane. Black circles in both main and inset panels show the number density of separations found using a 3" nearest neighbour match, for \textit{WISE} objects $N = 0.256-0.266\,\mathrm{mag}^{-1}\,\mathrm{deg}^{-2}$, $W1 = 15.47-15.03$, $\sigma_\mathrm{\alpha} = 0.091-0.151"$. Solid red lines show the full empirical AUF for these parameters. Dashed red lines show the empirical AUF without the inclusion of the differential source count breaks (Section \ref{sec:empaufzmdependence}), resulting in a distribution with larger perturbation offsets than seen in the data. Dotted red line in the inset panel shows a purely Gaussian AUF, represented by a Rayleigh distribution with uncertainty $0.121"$, entirely incompatible with the distribution of \textit{Gaia}-\textit{WISE} separations.}
    \label{fig:W1fitzmbreak}
\end{figure}

\begin{figure*}
    \centering
    \includegraphics[width=\textwidth]{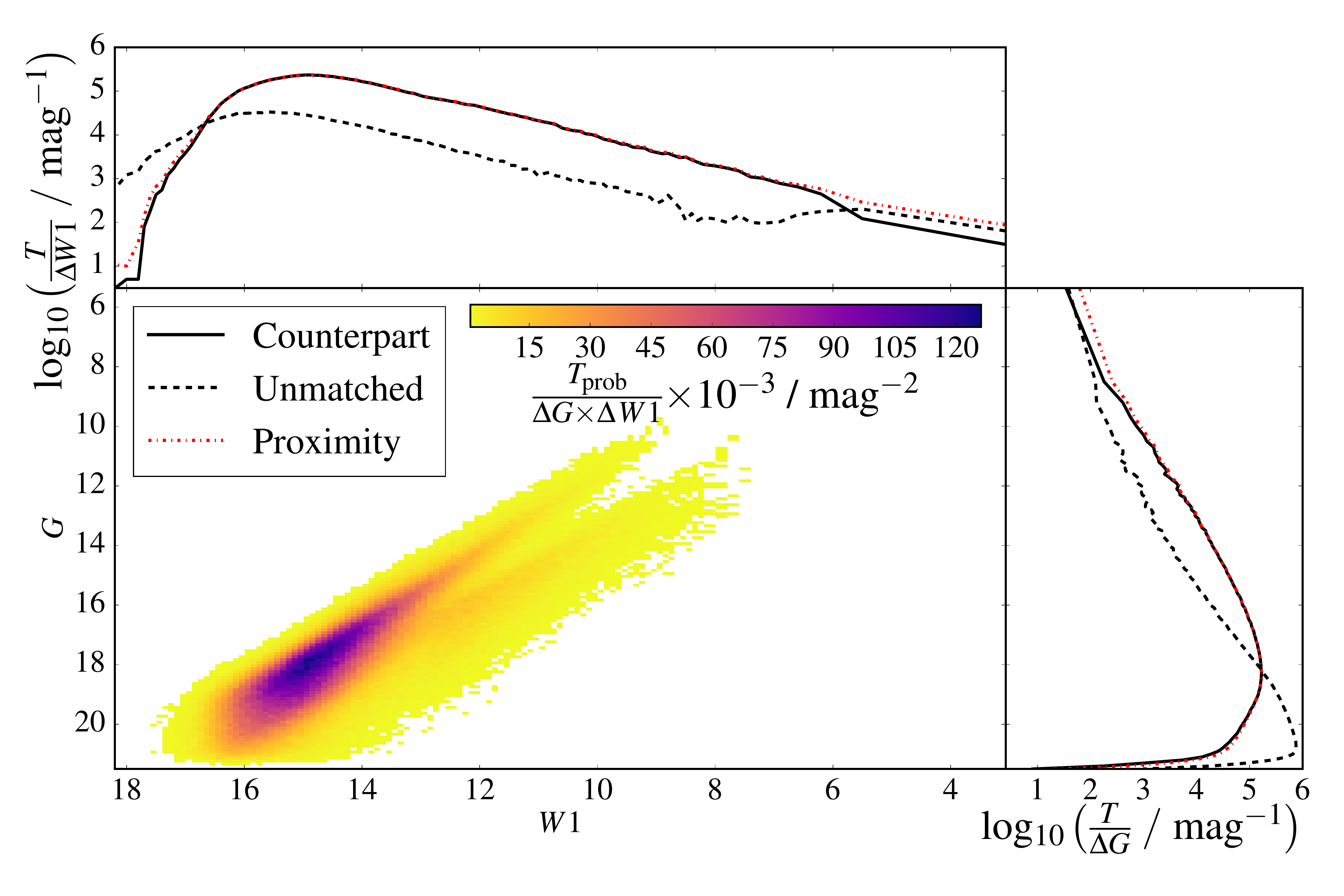}
    \caption{The number density of matched objects between \textit{Gaia} and \textit{WISE} using probability-based matching that includes the effect of crowding in the AUF for a 42 square degree region of the Galactic plane. Figure layout and colourbar are the same as Figure \ref{fig:gwgauss}. The empirical \textit{WISE} AUF results in a much more complete counterpart return rate, recovering more counterparts than the nearest neighbour-based match at $G \simeq 18$. It still rejects faint matches $G \geq 20$ as required.}
    \label{fig:gwemp}
\end{figure*}

An example of the full empirical AUF treatment is shown in Figure \ref{fig:W1fitzmbreak} (solid red line), compared to the 3" nearest neighbour matching of \textit{Gaia} and \textit{WISE} objects with $N = 0.261\,\mathrm{mag}^{-1}\,\mathrm{deg}^{-2}$, $W1 = 15.5$, and $\sigma_\mathrm{\alpha} = 0.121"$. The purely Gaussian AUF (dotted red line, inset panel) is completely incompatible with the separations seen in the data, and, indeed, highlights the main cause of the 45\% counterpart loss rate when using a Gaussian AUF (as discussed in Section \ref{sec:gaussaufeffect}). There is good agreement between the empirical AUF and the distribution of separations, however. The slight discrepancies between the separations and empirical distribution can be explained by a combination of the slight spreads in values of $N$, $W1$, and $\sigma_\alpha$ used to build the \textit{Gaia}-\textit{WISE} separations. Additionally, our treatment the effects of proper motions in our empirical AUFs could be incomplete, primarily suffering from incorrect epoch differences, causing us to miss a small additional source of perturbation seen in the separations between sources. We will discuss the inclusion of the effects of proper motions in AUFs further in Section \ref{sec:extendaufperturb}. However, our empirical AUF matches the distribution of source separations to high accuracy, in contrast with a pure Gaussian AUF (dotted line, inset panel, Figure \ref{fig:W1fitzmbreak}) which simply cannot explain the significant fraction of nearest neighbour matches at large separations.

\subsection{Empirical AUF Fitting Summary}
\label{sec:empaufsummary}
We can summarise the steps required to compute a given empirical AUF, including the effects of perturbation due to crowding, for a specific star as follows.
\begin{enumerate}
\item{Determine $N$, $m$ and $\sigma_\mathrm{pure}$.}
\item{Create a parameterisation of the differential source magnitude counts for the filter in question.}
\item{Assign random positions in the PSF to stars for a small magnitude offset range, drawing the number of stars according to Poissonian distribution.}
\item{Repeat the drawing of stars from the probability distribution for all magnitude offsets, accounting for differential source count variations with magnitude.}
\item{Using all contaminating stars within the PSF, determine the flux-weighted star position, to find the perturbation offset.}
\item{Repeat the perturbation offset calculation for a large number of PSFs, creating the offset distribution.}
\item{Convolve the offset distribution with a pure Gaussian of given uncertainty.}
\end{enumerate}

\subsection{The Effects of the Empirical AUF on \textit{Gaia}-\textit{WISE} Matches}
\label{sec:empaufeffect}
Now that we have constructed empirical AUFs, we can apply them to the same sky region as in Section \ref{sec:gaussauf}. While we still match between our two photometric catalogues using the method laid out by \citet{2018MNRAS.473.5570W}, we use our empirically constructed AUFs to define $G$. We also define our island cutoff radii, as well as counterpart and ``field'' star cut out radii as described by \citet{2018MNRAS.473.5570W}, using the new empirical AUFs. Assuming circular symmetry for the AUFs (see Section \ref{sec:circsymmetryauf}) simplifies the definition of $\mathcal{R}_Y$ somewhat, however, and we can now define it as 

\begin{equation}
\int\limits_0^{\mathcal{R}_Y}\int\limits_0^{2\pi}\! r\,G(r, \theta)\mathrm{d}\theta\,\mathrm{d}r = 2\pi \times \int\limits_0^{\mathcal{R}_Y}\! r\,G(r)\,\mathrm{d}r = Y.
\label{eq:raddist}
\end{equation}
We are more lenient than \citet{2018MNRAS.473.5570W} with our maximum offset due to the long, non-Gaussian tails, using the largest $\mathcal{R}_{0.99}$ of all \textit{WISE} stars in the matching region in question, slightly less complete than as with a Gaussian $G$. This slightly lower integral limit is still over an order of magnitude higher than that used in Section \ref{sec:gaussaufeffect}, due to the large effect contamination has on the \textit{WISE} positions. The nature of the non-Gaussian tails to the AUF mean that we must now cut our integrals at a slightly lower percentile than previously; see Section \ref{sec:simulatedfraction} for discussion of the effect this has on the matches obtained.

Matching the same catalogues as described in Section \ref{sec:gaussaufeffect}, the results of using the new PDF for our $G$ are shown in Figure \ref{fig:gwemp}, again accepting only matches with $P \geq 0.5$. We now recover the vast majority of our nearest neighbour-based counterparts. We also see a reduction of the number of faint (\textit{Gaia} $G \geq 20$, bottom of right inset panel) counterparts, when compared with the nearest neighbour matches, as expected. However, the objects recovered and rejected at the varying brightnesses in both the \textit{Gaia} and \textit{WISE} passbands require more detailed examination.

\begin{figure}
    \centering
    \includegraphics[width=\columnwidth]{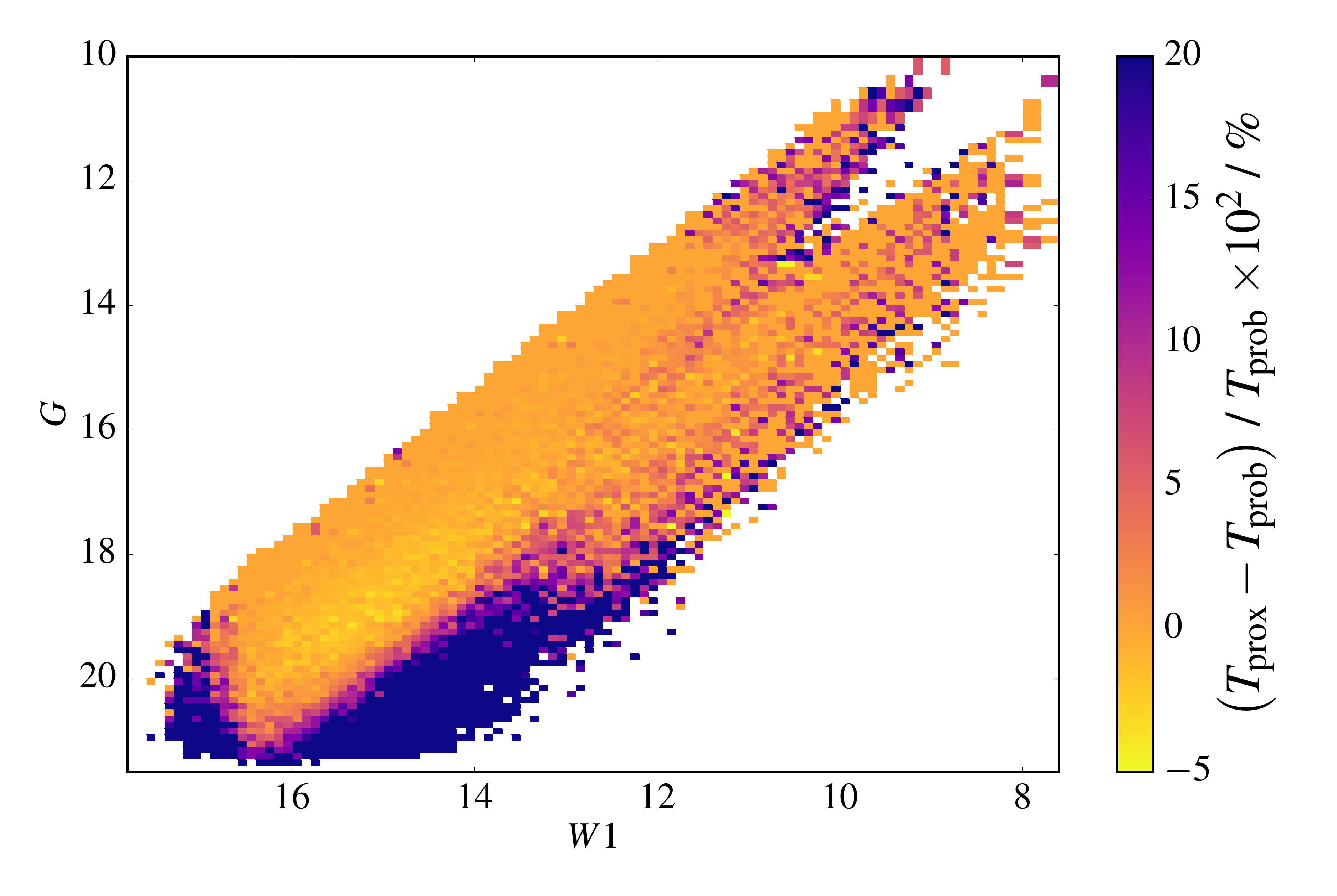}
    \caption{The relative difference in the number of objects in a 42 square degree region of the Galactic plane for \textit{Gaia}-\textit{WISE} objects. The magnitude density criterion is the same as for the main panel of Figure \ref{fig:gwgauss}. However, the colourbar shows the relative difference in probability- and nearest neighbour-based (``proximity'') matches. $W1 \leq 15$ there is a constant rejection of a small number of objects in all bins on the order of several percent, consistent with false match chance arguments. However, at $W1 \simeq 15$ there are two areas of importance. First, at $G\simeq18$ the probability-based matches return additional matches not picked up at a 3" nearest neighbour match, suggesting a small number of objects are astrometrically perturbed by $>$3". Second, at $G\simeq20$, there is a significant decrease in the number of matches.}
    \label{fig:gwempdiff}
\end{figure}

We therefore now consider the number of objects gained or lost by the probability-based matching process relative to the 3" nearest neighbour match, as shown in Figure \ref{fig:gwempdiff}. The first point of interest is that over much of the area occupied by bright ($W1 \leq 15$) matches there is a rejection of approximately 1-5\% of the matches, similar to the number of false positives (see Section \ref{sec:gaussaufeffect}). This indicates that our new AUF is still rejecting false matches, as expected. 

At faint magnitudes ($W1\simeq15$) there are two distinct regions of the magnitude-magnitude space. The first, at $G\simeq18$, is an area where extra pairings are picked up by the probability-based matching, which were not picked up by our nearest neighbour match. These are most likely objects which were astrometrically perturbed beyond our nearest neighbour cutoff radius, and therefore unable to be paired in the nearest neighbour match. The contamination at this magnitude is most likely to cause astrometric shifts which result in separations between \textit{Gaia} and \textit{WISE} source detections beyond the 3" nearest neighbour match radius \citep{2017MNRAS.468.2517W}. However, some of them could also be objects where the pair most favourable was not the closest. These objects would favour brighter, but further away, matches rather than some fainter, but closer, stars. This can be caused either by the brighter source having a larger absolute distance but smaller Mahalanobis distance, due to its smaller astrometric uncertainties, or by the photometric counterpart likelihood favouring the bright source over the faint object. The second region of interest, at fainter \textit{Gaia} magnitudes ($G\simeq20$), sees a loss of matches compared with nearest neighbour match for the same \textit{WISE} brightness ($W1\simeq15$). These could be the rejected faint nearest neighbour matches for the additional probability-based matches seen at $G\simeq18$. However, a fraction of these lost, faint \textit{Gaia} matches are \textit{WISE} objects which should match to \textit{Gaia} objects below the sensitivity level of the survey, which are coincidentally near to these objects of $G\simeq20$ whose corresponding \textit{WISE} object was removed from our catalogue in the process of cleaning poor quality data (see Table \ref{tab:flags}). This issue with incomplete datasets and quality selection can also explain the lack of bright pairings, where objects of $W1 \simeq 7$ should match \textit{Gaia} sources of $G \simeq 11$. Those rejected pairings (dashed lines, inset Figure \ref{fig:gwemp} panels) are primarly caused by saturation effects, with \textit{WISE} having a saturation magnitude $W1 \simeq 8$.

\begin{figure}
    \centering
    \includegraphics[width=\columnwidth]{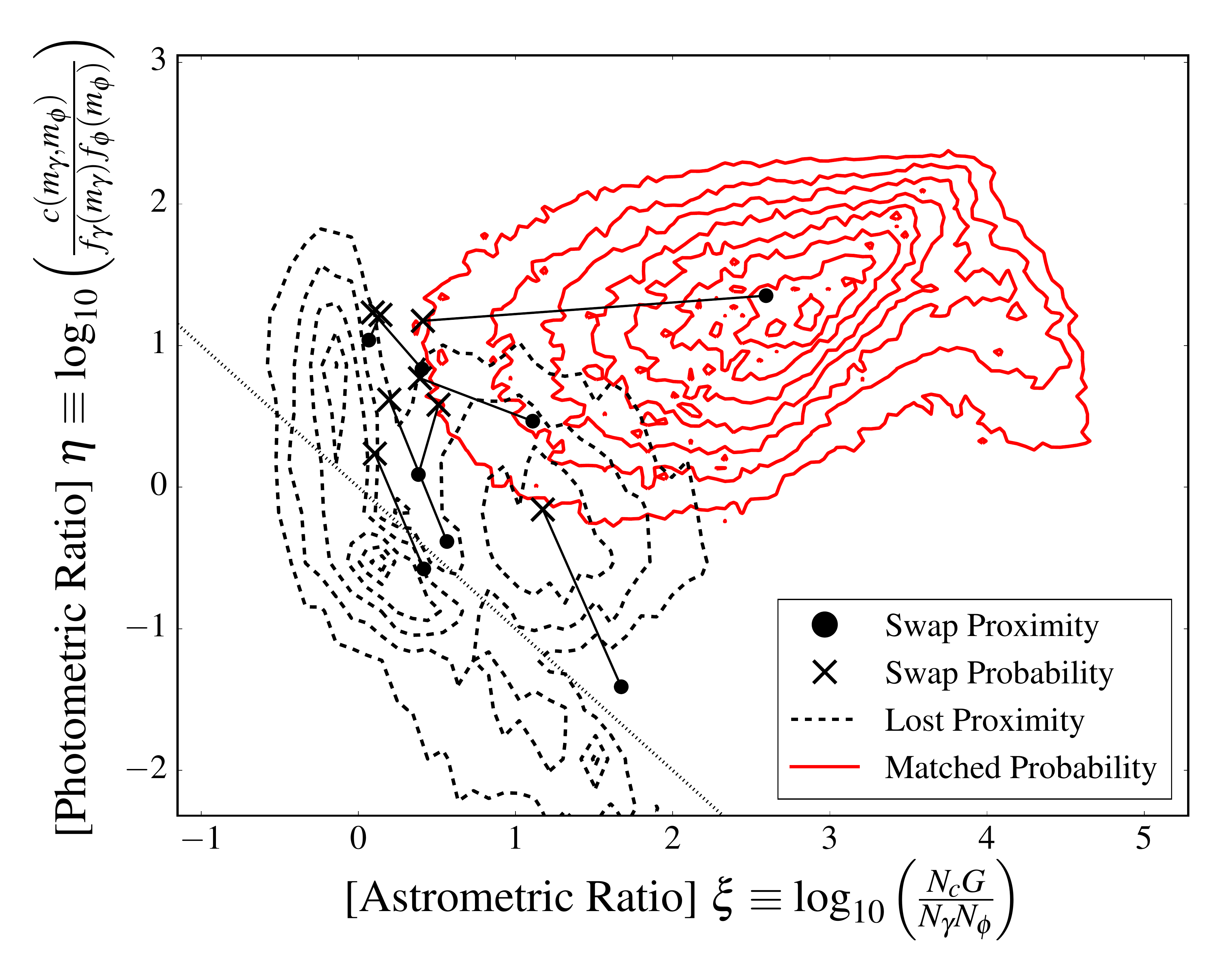}
    \caption{The astrometric and photometric likelihood ratios for \textit{Gaia}-\textit{WISE} matches for a 42 square degree region of the Galactic plane. The details of the figure are the same as Figure \ref{fig:likeratiogauss}, with the addition of crosses and filled circles connected by a solid black line. These represent the likelihood ratios of \textit{Gaia} objects which were nearest neighbour matched to one \textit{WISE} object (circles) but matched to a different star through the probability-based matching process (crosses). Those objects that were nearest neighbour matched but unmatched in a probability-based match lie at slightly higher than equal chance astrometrically, but are unlikely enough photometrically to drop below a combined equal likelihood of $\xi + \eta = 0$. These matches are therefore rejected by their mismatched photometry, rather than their spatial correlation (or lack thereof). In addition, almost all of the objects which swap their returned match shown increased photometric likelihood, indicating a more likely match based on their magnitudes in the two filters.}
    \label{fig:likeratioemp}
\end{figure}

We can analyse the acceptance and rejection of our nearest neighbour matches on probabilistic grounds by considering the likelihood ratios once more, shown in Figure \ref{fig:likeratioemp}. Most matches are still several orders of magnitude more likely matches than non-matches, based on their astrometry. Additionally, the spread of $\eta$ values ($\overline{\eta}\simeq1$, $0\lesssim\eta\lesssim2$) is the consistent with the case where the AUF was purely Gaussian. The differences arise when considering those objects rejected as probability-based matches which were nearest neighbour matched at 3". With the empirical $G$ term, the matches which are now lost with respect to the nearest neighbour matches still have $\xi\geq0$, but are an order of magnitude less likely to be a match to their nearest neighbour, than to be unrelated, photometrically (i.e., $\eta \simeq -1$). This suggests that those objects still not matched to a star positionally close to them when using the empirical AUF are rejected for flux-related reasons. This is in contrast to the Gaussian AUF case (cf. Figure \ref{fig:likeratiogauss}), where the losses were almost all astrometric. The inclusion of the photometric information for the use of empirical AUFs which have much larger non-Gaussian tails, on the order of several arcseconds, while simultaneously rejecting unphysical matches. Without the extra information provided by the magnitudes of the sources, our false match rate could potentially rise to unacceptable levels, resulting in untrustworthy merged datasets.

\begin{table*}
\centering
\begin{tabular}{c c l}
\hline
Column Name & FITS Name & Column Description\\
\hline
\textit{Gaia} Name & GAIA\_ID & ID of \textit{Gaia} source, as given by the \textit{Gaia} DR2 catalogue.\\
\textit{WISE} Name & WISE\_ID & ID of \textit{WISE} object, as given by the AllWISE catalogue.\\
\textit{Gaia} Right Ascension & GAIA\_RA & Right Ascension of \textit{Gaia} source, as given by the \textit{Gaia} DR2 catalogue.\\
\textit{Gaia} Declination & GAIA\_DEC & Declination of \textit{Gaia} source, as given by the \textit{Gaia} DR2 catalogue.\\
\textit{WISE} Right Ascension & WISE\_RA & Right Ascension of \textit{WISE} source, as given by the AllWISE catalogue.\\
\textit{WISE} Declination & WISE\_DEC & Declination of \textit{WISE} source, as given by the AllWISE catalogue.\\
$G$ & G\_MAG &\textit{Gaia} magnitude, as given by the \textit{Gaia} DR2 catalogue.\\
$W1$ & W1\_MAG & \textit{WISE} \texttt{w1mpro} magnitude, as given by the AllWISE catalogue.\\
$W2$ & W2\_MAG & \textit{WISE} \texttt{w2mpro} magnitude, as given by the AllWISE catalogue.\\
$W3$ & W3\_MAG & \textit{WISE} \texttt{w3mpro} magnitude, as given by the AllWISE catalogue.\\
$W4$ & W4\_MAG & \textit{WISE} \texttt{w4mpro} magnitude, as given by the AllWISE catalogue.\\
Match Probability & MATCH\_P & Overall probability that the \textit{Gaia} and \textit{WISE} sources are detections \\
& & of the same object, as given by equation 26 of \citet{2018MNRAS.473.5570W}.\\
$\eta$ & ETA & Photometric logarithmic likelihood ratio of \textit{Gaia}-\textit{WISE} match, \\
 & & as described by equation 37 of \citet{2018MNRAS.473.5570W}.\\
$\xi$ & XI & Astrometric logarithmic likelihood ratio of \textit{Gaia}-\textit{WISE} match, \\
& & as described by equation 38 of \citet{2018MNRAS.473.5570W}.\\
\textit{WISE} 1\% Contamination Probability & CONT\_P1 & Probability of \textit{WISE} source having contaminant of at least 1\% relative flux given \\
 & & its separation from its corresponding \textit{Gaia} detection, as given by equation \ref{eq:contamprob}.\\
 \textit{WISE} 10\% Contamination Probability & CONT\_P10 & Probability of \textit{WISE} source having contaminant of at least 10\% relative \\
 & & flux given its separation from its corresponding \textit{Gaia} detection.\\
Average \textit{WISE} Contamination & AVG\_CONT & Mean contaminating relative flux within PSFs when constructing $h_\mathrm{offsets}$ for \\
 & & \textit{WISE} sources at the local density, magnitude and intrinsic positional \\
 & & uncertainty of the primary source.\\
\hline
\end{tabular}
\caption{Table showing a description of column headers for the table. The sky position and magnitudes of the paired source in each respective catalogue is given to aid in the identification of sources. If further information, such as magnitude uncertainty, parallax, or proper motions, is required, then a match to the main dataset by source ID is recommended. The \textit{Gaia} DR2 data are available at \url{https://gea.esac.esa.int/archive/} and the \textit{WISE} AllWISE Source Catalog dataset is available at \url{https://irsa.ipac.caltech.edu/cgi-bin/Gator/nph-scan?submit=Select&projshort=WISE}. The full table is available at CDS via anonymous ftp to cdsarc.u-strasbg.fr (130.79.128.5) or via \url{http://cdsarc.u-strasbg.fr/viz-bin/qcat?IV/35}, and the University of Exeter ORE repository via \url{https://doi.org/10.24378/exe.629}.}
\label{tab:headers}
\end{table*}

We also have some cases where stars have swapped match between the nearest neighbour- and probability-based matches. These matches increase in $\eta$, suggesting a more likely match photometrically, possibly at the expense of a small amount of astrometric likelihood. Pairings which decrease in combined likelihood ratio (i.e., $\eta+\xi$) can be explained by the fact that these ratios consider the two stars in isolation. The full matching process considers all objects in both catalogues that are spatially correlated at once. This suggests that while the new probability-based match is slightly less favourable, another match considered jointly was more favourable, overcoming the slight loss in isolated likelihood ratio.

\section{Galactic Plane Matches}
\label{sec:gplanematch}
We provide here a list of \textit{Gaia} DR2-\textit{WISE} matches, available at CDS (anonymous ftp to cdsarc.u-strasbg.fr (130.79.128.5) or \url{http://cdsarc.u-strasbg.fr/viz-bin/qcat?IV/35}), and the University of Exeter ORE (\url{https://doi.org/10.24378/exe.629}). We follow the probability-based matching process of \citet{2018MNRAS.473.5570W}, with the addition of the construction of empirical AUFs as detailed in Section \ref{sec:empauf}. We also account for systematic proper motion drift between the two datasets, extrapolating the \textit{Gaia} positions to the \textit{WISE} epoch using the \textit{Gaia} DR2 proper motions where available. This catalogue contains all \textit{WISE} matches returned for \textit{Gaia} objects with $\lvert b \rvert \leq 10$. Therefore, if an object in either catalogue does not appear, it was returned unmatched, or did not meet our catalogue cleaning criteria, shown in Table \ref{tab:flags}. Additionally, it should be repeated that a not insignificant fraction of \textit{Gaia} objects will not have a detected \textit{WISE} counterpart due to their being merged inside a brighter \textit{WISE} object's PSF. Therefore, the non-matching of a \textit{Gaia} object should not necessarily be seen as an upper limit on the \textit{WISE} fluxes. 

It should be noted that we do not include matches with \textit{Gaia} astrometry in the inner region of the Galactic centre ($\lvert l \lvert\ \leq 10$, $\lvert b \lvert\ \leq 5$). In this region the extreme crowding causes such high flux contamination levels that the assumptions made about the source counts as a function of magnitude no longer hold, and any matches would be unreliable accordingly; see Section \ref{sec:extremecrowding} for more details. We also only clean the \textit{Gaia} dataset to the lowest of suggested levels, simply removing sources included in the \textit{Gaia} DR2 catalogue that show strong signs of being non-physical due to their poor unit weight error criterion (see Table \ref{tab:flags} and appendix C of \citealp{Lindegren2018} for further details). When using the catalogue, we recommend that more stringent selection criteria are applied to ensure that only high quality matches are utilised. We provide this catalogue of \textit{Gaia}-\textit{WISE} matches as a general composite dataset to be further refined based on the individual needs of each science case.

We also caution the reader to not accept the non-matching of the very brightest sources ($G \lesssim 10$, $W1 \lesssim 8$) without individual confirmation. The reasons for the rejection of these expected matches are three-fold. These sources are expected to exhibit significant proper motions, and our assumed five-year baseline between observations may not apply to all observations. The \textit{WISE} dataset is a composite of the original All-Sky mission and a reactivated NEOWISE mission \citep{2014ApJ...792...30M}, four years after the initial mission was successfully completed. With the extremely high precision positions of these bright sources, even objects with proper motions as low as $50\mu \mathrm{as}\, \mathrm{year}^{-1}$ would be assumed positionally uncorrelated if the \textit{WISE} epoch were incorrect when correcting for the proper motion positional systematic. In addition, the ``DOF Bug'' (appendix A, \citealp{Lindegren2018}) resulted in under-estimated formal uncertainties for \textit{Gaia} stars $G\leq13$. Finally, these bright sources are in the saturated regime of $W1-3$ (with $W1$ saturating around 8$^\mathrm{th}$ magnitude), and thus the positions of these sources are subject to additional systematics due to the nature of deriving positions and corresponding uncertainties from non-saturated PSF wings. We therefore recommend for these brightest sources a naive nearest-neighbour match to confirm the presence of a similarly bright source in the opposing catalogue.

To provide a useful catalogue of matches, we not only provide the source pairings and some key information (e.g., positions, magnitudes, names), but additional information to allow the user to evaluate whether they wish to accept the pairing. We follow the ``Full Coverage'' method outlined by \citet{2017MNRAS.468.2517W} in their section 9.2. When accepting a source pairing, here accepting the most likely match hypothesis without regard to its value in contrast with the discussion above, we also calculate the probability of the \textit{WISE} source being contaminated by blended objects. Table \ref{tab:headers} describes the data columns for the match catalogue: \textit{Gaia} and \textit{WISE} object names, \textit{Gaia} and \textit{WISE} astrometric positions ($\alpha$, $\delta$), \textit{Gaia} $G$ magnitude, \textit{WISE} magnitudes, and probability of set the pair was matched in (see \citealp{2018MNRAS.473.5570W} for details). The final columns are the probability that the \textit{WISE} detection in the matched pair is a contaminated detection, as given by equation \ref{eq:contamprob}, and the respective average \textit{WISE} contamination flux ratio for an empirical AUF of the appropriate local density normalisation, magnitude and instrinsic astrometric uncertainty.

\subsection{Source Contamination Probability}
\label{sec:contamprob}
At a given separation, the probability of a match being contaminated by an additional source of flux, denoting this hypothesis as $\psi$, is

\begin{equation}
P\left(\psi | r\right) = \frac{P\left(\psi\right)p\left(r | \psi\right)}{p\left(r\right)},
\label{eq:contamprob_simple}
\end{equation}
where $r$ is the separation between the two matched sources. For a two-directional match this equation is slightly more complex, considering the hypotheses that both objects are contaminated, one but not the other source is contaminated, and the chance that neither object is affected by systematic perturbations. We can consider each source in turn, representing the hypotheses that a \textit{Gaia} source is contaminated as $\psi$ and uncontaminated as $\widetilde{\psi}$, respectively. Analagously, the hypotheses are $\omega$ and $\widetilde{\omega}$ for the cases of a contaminated and uncontaminated \textit{WISE} source respectively. Therefore, if we wish to consider the hypothesis that a given source is contaminated given the separation between it and its corresponding detection, denoted $P_\mathrm{contam}$ henceforth, we can marginalise over both hypotheses for the match in the opposing catalogue. This would give

\begin{align}
\begin{split}
P\left(\omega | r\right) &= P\left(\omega, \psi | r\right) + P\left(\omega, \widetilde{\psi} | r\right) \\&= \frac{P\left(\omega\right)P\left(\psi\right)p\left(r | \omega, \psi\right) + P\left(\omega\right)P\left(\widetilde{\psi}\right)p\left(r | \omega, \widetilde{\psi}\right)}{p\left(r\right)},
\label{eq:contamprob}
\end{split}
\end{align}
for the hypothesis of the \textit{WISE} source being contaminated, assuming the priors for each catalogue suffering contamination are independent from one another. The evidence is given by the combination of all four hypotheses,

\begin{align}
\begin{split}
p(r) =\ &p\left(r | \psi, \omega\right)P\left(\psi\right)P\left(\omega\right) + p\left(r | \psi, \widetilde{\omega}\right)P\left(\psi\right)P\left(\widetilde{\omega}\right) + \\&p\left(r | \widetilde{\psi}, \omega\right)P\left(\widetilde{\psi}\right)P\left(\omega\right) + p\left(r | \widetilde{\psi}, \widetilde{\omega}\right)P\left(\widetilde{\psi}\right)P\left(\widetilde{\omega}\right).
\label{eq:contamprobnorm}
\end{split}
\end{align}

The priors for the contamination hypotheses are simply the fraction of numerical PSF simulations (Section \ref{sec:constructempauf}) to suffer from additional sources with a total flux ratio greater than 1\% for each catalogue in turn. To evaluate the likelihood of each joint hypothesis, the evaluation of the convolution of $h_\mathrm{pure}$ for both catalogues and $h_\mathrm{offsets}$ for any catalogue in which the contamination hypothesis is being considered is required. For the case above of a contaminated \textit{WISE} source and uncontaminated \textit{Gaia} detection, our likelihood is

\begin{equation}
p\left(r | \omega, \widetilde{\psi}\right) = \left(h_{\omega, \mathrm{pure}} * h_{\psi, \mathrm{pure}} * h_{\omega, \mathrm{offsets}}\right)\big(r\big),
\label{eq:likelihoodexample}
\end{equation}
where the syntax $(f*g)(x)$ represents the convolution of functions $f$ and $g$ evaluated at $x$. We include these probabilities in the \textit{Gaia}-\textit{WISE} composite catalogue to aid in the selection of uncontaminated \textit{WISE} sources, with the \textit{Gaia} probability provided for symmetry and completeness. Wherever used, $P_\mathrm{contam}$ represents the probability that the source in the given catalogue in question suffers contamination above 1\% relative flux given the separation between it and its corresponding detection, independent of potential contamination in the detection in the opposing catalogue.

\subsection{Galactic Plane Match Testing}
\label{sec:matchtests}
In Section \ref{sec:empaufeffect} we analysed a representative region of the Galactic plane, comparing our improved empirical AUF treatment to a naive nearest neighbour match and a simplistic, pure Gaussian AUF. In this subsection we examine our matching process in more detail, discussing a variety of tests applied to the \textit{Gaia}-\textit{WISE} matches.

\subsubsection{The Effect of Simulated Source Counts on Match Fractions}
\label{sec:simulatedfraction}
The first test we examine is that of the effect of the simulated AUFs on the pairings obtained. Both the creation of the perturbed distribution (Section \ref{sec:constructempauf}) and the formulation of the differential source counts used to evaluate PSF circle densities (Section \ref{sec:empaufzmdependence}) use stochastic processes, and therefore will change with each iteration. To quantify the level of variation these stochastic processes introduce, we ran two identical matches on the catalogue used in Sections \ref{sec:gaussauf} and \ref{sec:empauf}. Of the $\approx675000$ matches in the region, approximately 200 pairings were not shared by both composite catalogues, on the order of 0.035\% of matches. The acceptance or rejection of these matches lies in equation \ref{eq:raddist}. Depending on the subtle variations in the empirical AUF created in each match process, these sources lie either just inside, or just outside of, the 99$^{\mathrm{th}}$ percentile of the AUF integral. They are therefore rejected in one run as being incompatible astrometrically, but accepted in the other. This effect is an unavoidable side effect of using empirical treatments, and should be considered carefully for cases where sources might be separated by large distances, such as high proper motion sources.

\begin{figure*}
    \centering
    \begin{subfigure}[b]{0.33\textwidth}
        \centering
        \includegraphics[width=\textwidth]{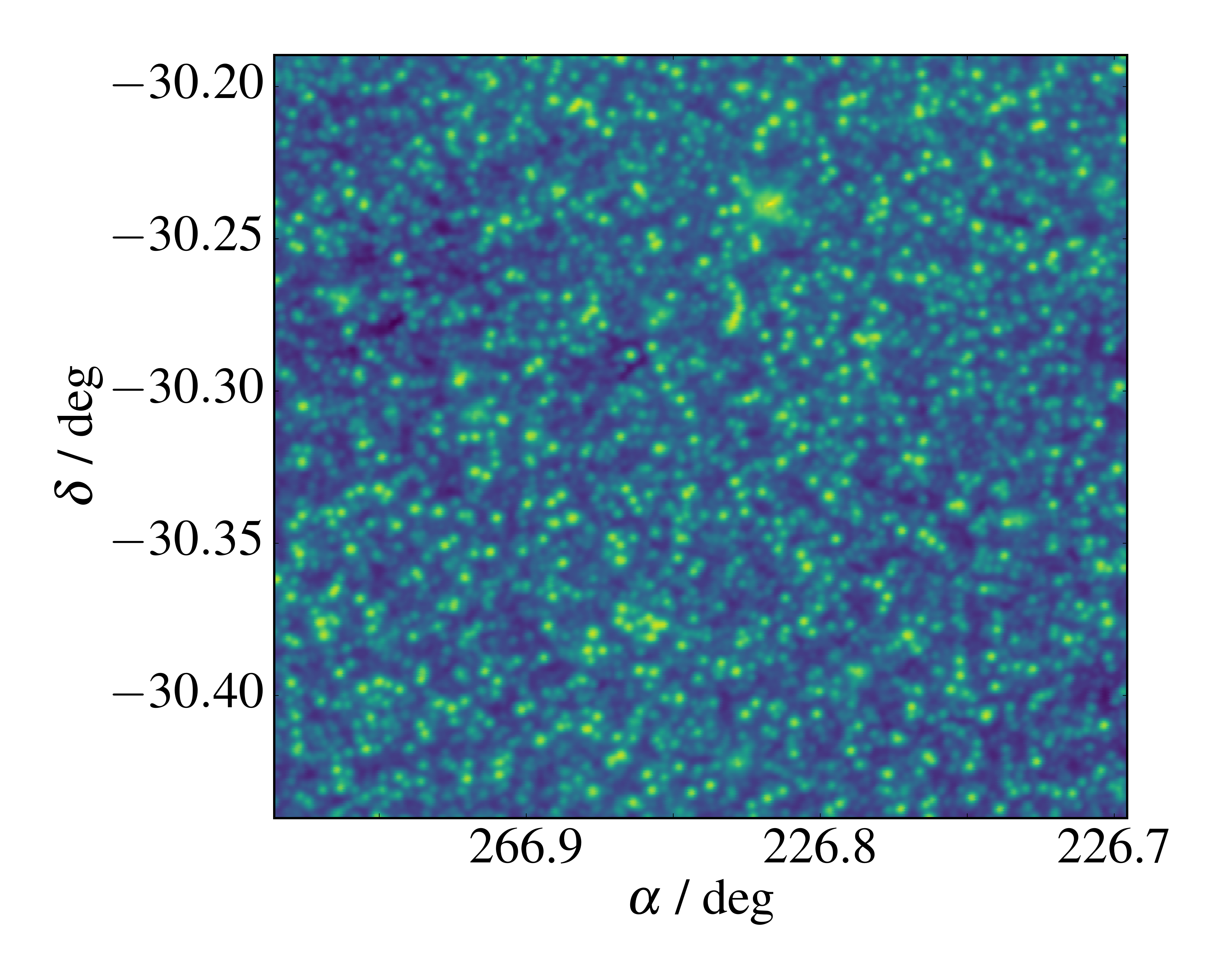}
        \caption{}  
        \label{fig:wiseview0}
    \end{subfigure}
    \begin{subfigure}[b]{0.33\textwidth}  
        \centering 
        \includegraphics[width=\textwidth]{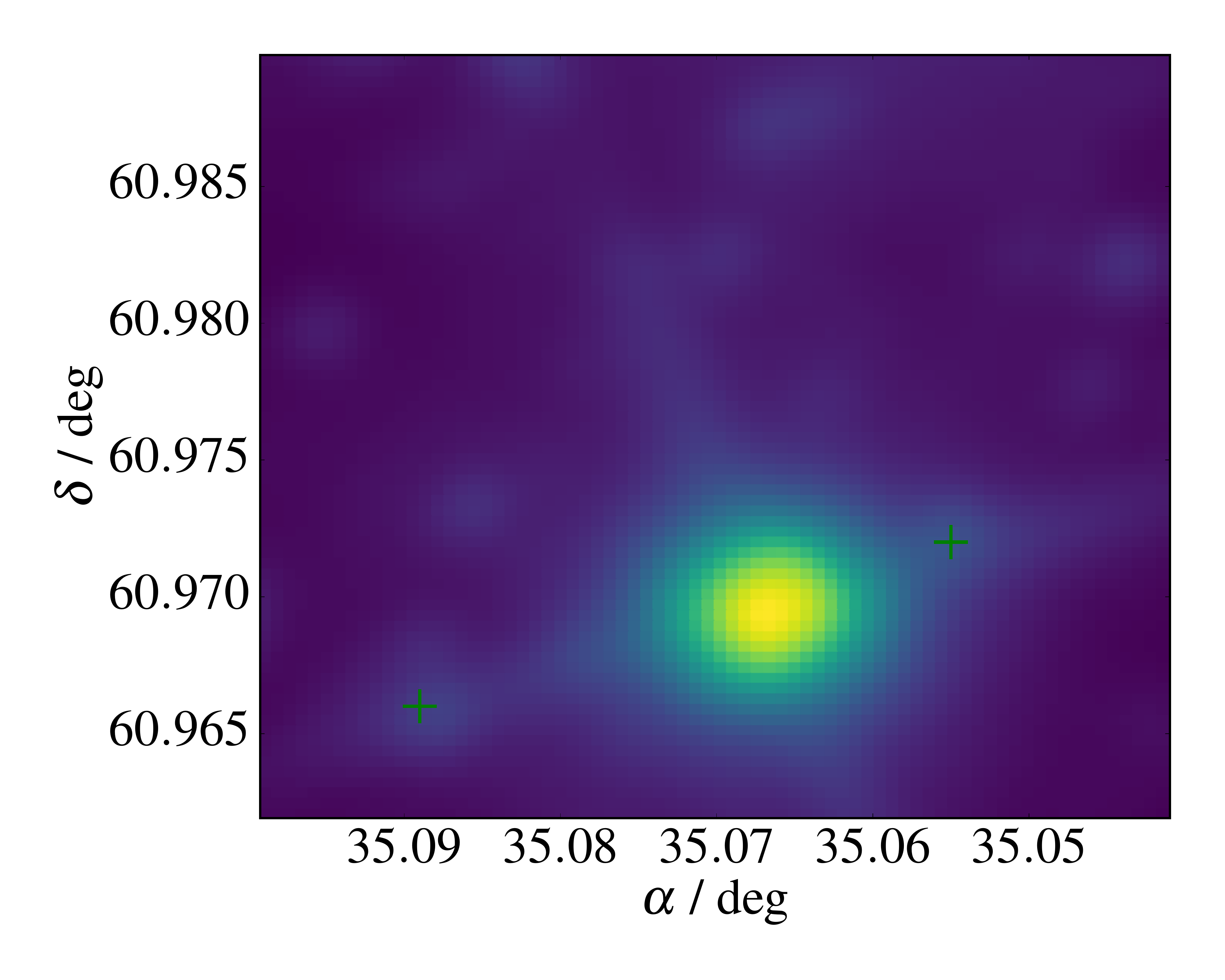}
        \caption{}   
        \label{fig:wiseview1}
    \end{subfigure}
    \begin{subfigure}[b]{0.33\textwidth}   
        \centering 
        \includegraphics[width=\textwidth]{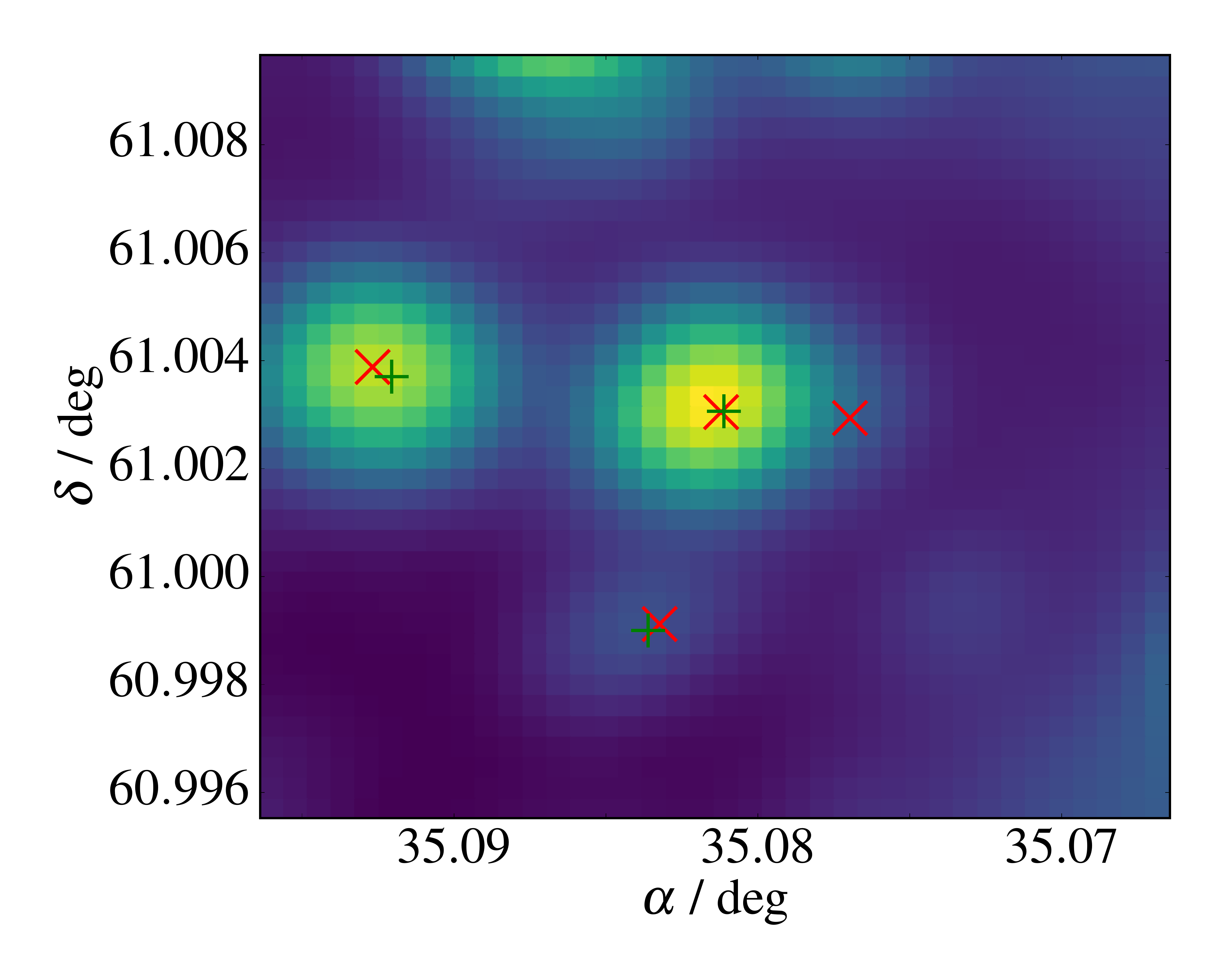}
        \caption{}    
        \label{fig:wiseview2}
    \end{subfigure}
    \caption{Three examples of the importance of follow-up visual inspection of survey images in the case of uncertain results. (a) A 0.0625 square degree view of the Galactic centre, at $l=359, b=-1$, highlighting the effects of extreme crowding (see Section \ref{sec:extremecrowding} for more details). (b) An artefact in an example \textit{WISE} image, for a field of view 100" on each side centred on $l=133.5, b=-0.06$. Diffraction spikes caused by a saturated source ($W1=6.7$) may affect the measured flux and position of the two \textit{WISE} sources (green plusses) resolved within the spikes on either side of the bright source. (c) An example of the failure to deblend sources in \textit{WISE} due to its large PSF. The image shows a field of view of 50" on each side, centred on $l=133.5, b=-0.03$. The green plusses show \textit{WISE} sources, with centroid positions lying on the peaks of the \textit{WISE} images. The red crosses show \textit{Gaia} detections, showing that the central \textit{WISE} source ($W1=13.9$) is resolved into two \textit{Gaia} sources ($G=15.6$ and $G=19.3$).}
    \label{fig:wiseview}
\end{figure*}

\subsubsection{The Effect of Normalisation Radius on Match Rate}
\label{sec:normalisationfraction}
Another potential source of variation in our matching process is the local density normalisation (equation \ref{eq:solveforn}). To evaluate the level our choice of normalisation radius affected our results, we ran a match identical that used in Section \ref{sec:empaufeffect}, but with a one degree normalisation radius, rather than 15 arcminutes as is used in all other cases. We found that there were on the order of 1000 matches differing between our two matches, similar in magnitude to the variation due to stochastic processes used in the matching process (Section \ref{sec:simulatedfraction}). This low level of match variation suggests that the density of sources does not vary significantly between one degree and 15 arcminute radii, and that our evaluation of $N$ for each source is robust. Small variations of $N$ do have a minor effect on matches near to the 99$^{th}$ percentile separation of their corresponding AUFs, however.

\subsubsection{Analysis of the \textit{Gaia}-\textit{WISE} False Match Rate}
\label{sec:falsematchrate}
A useful metric for consideration of any dataset is its false match rate. We can quantify this level by matching between a \textit{Gaia} catalogue from one region of the Galactic plane, and a \textit{WISE} catalogue from a second region, under the assumption that the positions of sources across the Galactic plane are independent from one another. To achieve this we took all \textit{Gaia} sources $121 \leq l \leq 128$, $-3 \leq b \leq 3$, filtered them for quality as per the criteria in Table \ref{tab:flags}, then incremented their Galactic longitude by 10 degrees (i.e., if a star is recorded at $l = 125$, $b=0$, we ``moved'' it to $l = 135$, $b=0$). We then ran a match between this new \textit{Gaia} catalogue and our original \textit{WISE} catalogue, returning $\approx5000$ matches, or 0.6\% of the input \textit{WISE} catalogue. This highlights the improvements the additional information available to our matching process, compared with a simple nearest neighbour match. A nearest neighbour match should return 4\% false matches (Section \ref{sec:gaussaufeffect}; \citealp{2017MNRAS.468.2517W}). We see almost a factor ten improvement in our false match rate, with both the variable scale length and inclusion of the photometric information allowing for the identification and rejection of 7 out of every 10 uncorrelated star pairs.

\subsubsection{The Effect of Photometric Likelihood Inclusion on Match Fraction}
\label{sec:photometryinclusionfraction}
We can examine further the effect the inclusion of the photometric information of the catalogues has on the matches returned by the matching process. If we remove from consideration the weighting of the hypotheses by star brightnesses we can analyse the pairings accepted and rejected, and the relative probabilities they are assigned. Setting $c(m, m) = f(m)f(m) = 1$ (see section 3 of \citealp{2018MNRAS.473.5570W} for more details), our matching process returned $\approx680000$ matches, cf. the $\approx675000$ matches obtained with the photometric probability densities' inclusion of which $\approx665000$ matches are shared between the two matching processes. As expected, the photometric likelihood ratio being included allows for the inclusion of $\simeq1\%$ of matches, but more crucially rejects 70\% of the $\simeq4\%$ of serendipitous matches expected to occur. Comparing the match probabilities for the common matches accepted by both processes the inclusion of the photometric information improves the overall probability of acceptance. If we compare the median Bayes' factor of the null hypothesis (that these two sources being unrelated detections), we find an increase of a factor of approximately ten (cf. the photometric likelihood ratio $\eta$, Figure \ref{fig:likeratioemp}). 

\subsection{Visual Inspection of Cross-Matches}
\label{sec:visualinspect}
The test matches in Section \ref{sec:matchtests} show that the matches we provide in Section \ref{sec:gplanematch} are good statistically, as a large-scale global population. However, there are cases in any cross-match procedure which require confirmation, especially if the numbers of sources in one or both of the catalogues are low. These could be caused by small-number statistics, cases of accidental catalogue entry duplication resulting in an unphysically low match probability, or decisions made during the catalogue creation process, either keeping noise artefacts as real data or removing physical sources from the resultant dataset. We highlight here a few cases that may merit closer inspection, as examples for consideration. The cross-matched catalogue we present in Section \ref{sec:gplanematch} is robust; however, the individual science case may require further examination. In any case where an individual cross-match does not follow the expected, ensemble trend -- such as in cases where an outburst has significantly affected the magnitude of a source between two catalogues at two different epochs -- or in cases where one catalogue only contains a small number of sources, we advise the reader to visually inspect the catalogue images to ascertain the veracity of questionable cross-matches.

Figure \ref{fig:wiseview} shows three examples of instances where the cross-matches of \textit{Gaia} sources to \textit{WISE} objects may not necessarily be trustworthy. Figure \ref{fig:wiseview0} shows a 0.25 degree by 0.25 degree view of the Galactic centre centred on $l=359, b=-1$. At such extreme on-sky densities, even the brightest sources are significantly crowded, leading to uncertainties in the ability of the \textit{WISE} pipeline to successfully deblend such sources. See Section \ref{sec:extremecrowding} for more discussion of the effects of extreme crowding.

Another example of an issue affecting photometric catalogues is that of artefacts in the survey images. One of the most significant examples of image artefacts is the diffraction spikes caused by extremely bright sources. Figure \ref{fig:wiseview1} shows such an example, showing a $W1=6.7$ source in a 100" by 100" \textit{WISE} image at $l=133.5, b=-0.06$. The large diffraction spikes may affect the determination of the position and brightnesses of the two nearby sources, shown in green plusses; see Section \ref{sec:wisedeblend} for more discussion of the deblending of sources in the \textit{WISE} pipeline.

Figure \ref{fig:wiseview2} shows a visual example of the blending of sources in the \textit{WISE} catalogue, in a 50" by 50" image centred on $l=133.5, b=-0.06$. Green plusses indicate the \textit{WISE} sources recorded in the \textit{WISE} catalogue, while red crosses show the \textit{Gaia} sources detected in the same sky region. Here the central \textit{WISE} source, $W1=13.9$, has been resolved into two \textit{Gaia} sources due to the superior \textit{Gaia} angular resolution, detected as one source with $G=15.6$ and a fainter source of $G=19.3$. This second \textit{Gaia} source is not deblended in the \textit{WISE} catalogue, resulting in an additional $\simeq3\%$ flux introduced into the blended \textit{WISE} source, and perturbing the recorded centroid towards the fainter object (seen at smaller right ascension).

\subsection{Contamination Level of Resulting Cross-Matches}
\label{sec:contaminationlevel}
\begin{figure}
    \centering
    \includegraphics[width=\columnwidth]{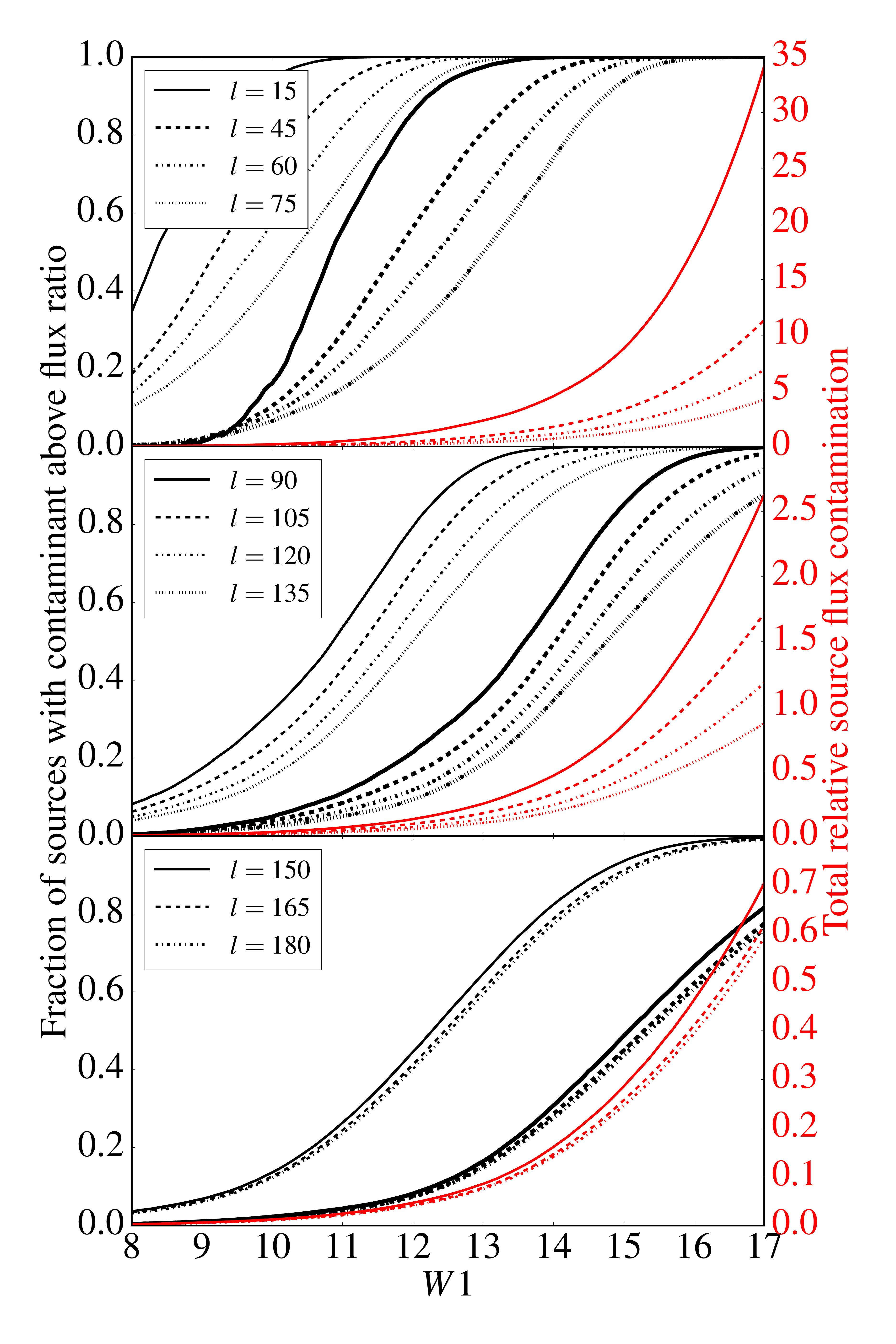}
    \caption{The level of contamination of simulated TRILEGAL sources, constructed following Sections \ref{sec:constructempauf} and \ref{sec:empaufoveralldependence}, for a range of Galactic longitudes $l=15-180$. The red lines show the average total flux, relative to the central source, of contaminants inside simulated PSFs of sources $W1=8-17$. The black lines show the fraction of simulated central sources with a contaminant of greater than 1\% and 10\% flux, relative to the central source, for thin and thick lines respectively.}
    \label{fig:wisecontamprob}
\end{figure}

Another useful metric when considering any pairings in the catalogue we present in this paper is the level of contamination suffered by the sources. \textit{Gaia}, with its relatively bright completeness limit and excellent angular resolution, is effectively uncrowded, suffering less than 0.1\% crowding \citep{2017MNRAS.468.2517W}. \textit{WISE}, however, suffers significant crowding. Figure \ref{fig:wisecontamprob} shows some simulated \textit{WISE} sources, using TRILEGAL to simulate small sections of the Galactic plane at $b=0$, from close to the Galactic centre ($l=15$), to the Galactic anti-centre ($l=180$).

For magnitudes in the range $W1=8$ to $W1=17$, a million test PSFs were simulated for the given central source brightness. Simulated contaminant sources were placed randomly within their PSFs according to the number density of sources, following the method laid out in Sections \ref{sec:constructempauf} and \ref{sec:empaufoveralldependence}, using the TRILEGAL differential source counts at the given Galactic longitude (cf. Figure \ref{fig:trilegalcounts}). The red lines in Figure \ref{fig:wisecontamprob} show, for each Galactic longitude, the average additional flux within the PSFs of a source as a function of its given central magnitude. This flux is approximately given by $\sum_{\Delta m} \left(P_B(\Delta m) \times -2.5 \log_{10}(\Delta m)\right)$ -- roughly given as the number of expected sources multiplied by their relative flux ratio, summed over all magnitude offsets. The black lines show the fraction of the test PSFs which contain a contaminating source with relative flux ratios above 1\% (thin black lines) or 10\% (thick black lines). This fraction, in the 1\% case, is the same as the prior $P(\omega)$ used in equation \ref{eq:contamprob}.

As can be seen in Figure \ref{fig:wisecontamprob}, there is an overall trend with Galactic longitude and thus overall source density, with lower typical source density leading to lower flux contamination levels, and a lower probability of contamination by a source above a given flux ratio, for a given central source brightness. At a fixed longitude, the contamination level and fraction increase as a function of decreasing central source brightness, as expected. However, the level of contamination in the \textit{WISE} dataset is severe. In a medium case, $l=135$ -- the region used to construct the tests described in Section \ref{sec:matchtests} -- sources suffer on the order of 25\% additional brightening at $W1=14$, and have a $\simeq35\%$ chance of perturbation by an individual source with at least 10\% the brightness of the central source, potentially leading to significant astrometric perturbation. Even at relatively bright magnitudes ($W1=10$) a source at $l=135, b=0$ has a 15\% chance of contamination by a blended object of at least 1\% relative flux.

This level of contamination means that almost all of the \textit{WISE} sources in this cross-match suffer some level of compromised position and flux measurements. We therefore do not recommend using the \textit{WISE} dataset in cases where additional flux would impact the scientific results obtained, such as when infrared (IR) excesses are important, or the photometry is being used to calculate a distance modulus. We provide individual average \textit{WISE} flux contamination and contamination probability for each \textit{WISE} source in the merged dataset (see Table \ref{tab:headers} for more details on these two columns).

\section{Discussion}
\label{sec:discussion}

\subsection{Deblending of \textit{WISE} Catalogue Sources}
\label{sec:wisedeblend}
Throughout this work we have assumed that the \textit{WISE} catalogue is subject to 100\% blending. However, the AllWISE data reduction process includes two stages to alleviate the effects of crowding: passive and active deblending. While this process may allow for some alleviation of crowded sources, most contaminants will remain blended, reducing the density of sources surrounding bright sources (see, for example, figure 3 of \citealp{2017MNRAS.468.2517W})

The passive deblend is a relatively simple process, whereby \textit{WISE} sources with overlapping PSF wings -- sources with centroids within 24" of one another -- have their brightnesses derived simultaneously. This simultaneous fitting ensures that the small extra flux that would otherwise be attributed to a given source is removed from its measured brightness and assigned to the nearby neighbouring source. However, passive deblending can only be done on sources up to 2.5 magnitudes fainter than the primary source in the blend group, allowing for the fitting of only those sources brighter than 10\% the flux of the primary object, with all other potential, fainter sources discarded at this stage.

Once the set of all overlapping sources have had positions and fluxes calculated, if the reduced $\chi^2$ statistic of the model fit to the data is above a minimum threshold, then the data are deemed unexplained by the model. There is assumed to be an unseen, unblended \textit{WISE} source, and an additional object is added to the model. However, only one additional source is ever added through active blending, and only if the reduced $\chi^2$ is above a critical threshold; in fact, the majority of sources in the AllWISE catalogue have zero additional blended sources, active or passive.

As the majority of sources in the AllWISE catalogue are still subject to any unseen contaminant objects, with few sources resolved during the catalogue creation, these must be handled through indirect means. The data reduction process does some allow for the potential of removing some of the potential contaminant sources blended with brighter \textit{WISE} objects. However, it does not resolve almost all of the blended \textit{WISE} sources; additionally the method laid out in Section \ref{sec:empauf}, with its implicit assumption of zero deblending, holds for the case of the \textit{WISE} catalogue (cf. Figure \ref{fig:W1fitzmbreak}), and thus the assumption of 100\% source blending is a reasonable one.

\subsection{Comparison with Literature Catalogue Matching Methods}
\label{sec:literaturecomparison}
It is useful at this point to compare our method, and the results we obtained, to those currently available in the literature. 

\subsubsection{Comparison with Pure Gaussian AUF Literature Matching Methods}
\label{sec:literaturegausscomp}
The most obvious difference between the method laid out here, building upon the probability-based matching processes laid out by \citet{2018MNRAS.473.5570W}, and previous literature works, is the effect of relaxing of the assumption of Gaussianity in the AUF. When using a pure Gaussian AUF we match 57\% of the sources returned with a fully empirical AUF that takes into account the effects of crowding. Therefore any cross-matching method that does not take this or any additional perturbations into account will underestimate its match fraction significantly. While \textit{WISE} is perhaps one of the more extreme cases for crowding, being a deep and complete survey with a large PSF, these effects are still non-negligible for other catalogues. For example, 2MASS \citep{Skrutskie:2006um} suffers crowding at its median magnitude that causes on the order of 10\% of stars to be perturbed beyond the separation where a Gaussian-only AUF would successfully recover them. Even as bright as $K_\mathrm{s}\simeq12$ this effect is at the 3\% level. 

It is therefore critical that these systematic effects -- perturbations due to crowding, but more generally any systematics such as proper motion, parallax, astrometric solution offsets, etc. -- are included in the AUFs of these catalogues. The general formalism of the AUF derived by \citet{2018MNRAS.473.5570W} allows for these effects to be folded in trivially; see Section \ref{sec:extendaufperturb} for more details. We therefore recommend the reader consider the catalogues being cross-matched, particularly with reference to the typical density, sources per PSF circle, before accepting the results of any cross-match involving a pure Gaussian AUF (e.g., \citealp{Sutherland:1992aa} and any work building upon their ``LR'' method, such as \citealp{Pineau:2017aa}; \citealp{Budavari:2008aa}; \citealp{2018MNRAS.473.4937S}; or \citealp{2017A&A...607A.105M}).

\subsubsection{Direct Match Comparison with \textit{Gaia} DR1}
\label{sec:gaiadr1comp}
We can perform a more direct comparison to a literature cross-match, comparing our matches to those provided as part of the \textit{Gaia} Data Release 1 (DR1; \citealp{2016A&A...595A...1G}; \citealp{Gaia-Collaboration:2016aa}; \citealp{2017A&A...607A.105M}). As part of the release they provide cross-matches between \textit{Gaia} and \textit{WISE}, allowing for an analysis of the matches between the two methods. Our method returns 86\% of all \textit{WISE} sources as being matched to a \textit{Gaia} DR1 source in the 42 square degrees of the Galactic plane centered on $l=135$, $b=0$, in good agreement with the official \textit{Gaia} DR1 match fraction (figure 3n, \citealp{2017A&A...607A.105M}). However, the extra matches they obtain, compared with the match rate we find in Section \ref{sec:gaussauf}, are a result of the broadening of their astrometric uncertainties (section 3.2 of \citealp{2017A&A...607A.105M}), which they believed accounted for epoch differences and any resultant proper motion shifts of the sources. These broadened astrometric uncertainties are much larger than the typical precision of either dataset, leading to the case where the parameters of the Gaussian AUF are independent of the properties of the sources themselves. The approximately constant uncertainties lead, effectively, to a reduction to a nearest neighbour match, with a matching radius that depends on the local source density. This radius, in most cases, is sufficiently large to capture the non-Gaussian wings of the full AUF, resulting in most pairings successfully being recovered. Their analytical solution is useful, allowing for simpler computations and the flexibility to include the relative likelihood of multiple matches. \citet{2017A&A...607A.105M} use this advantage to assign multiple \textit{Gaia} ``mates'' to singular \textit{WISE} counterparts, accounting for the higher \textit{Gaia} angular resolution deblending otherwise confused sources. However, the uncertainty broadening required to provide a good match rate, overcoming the astrometric perturbation from this crowding, has another, more subtle effect.

\begin{figure}
    \centering
    \includegraphics[width=\columnwidth]{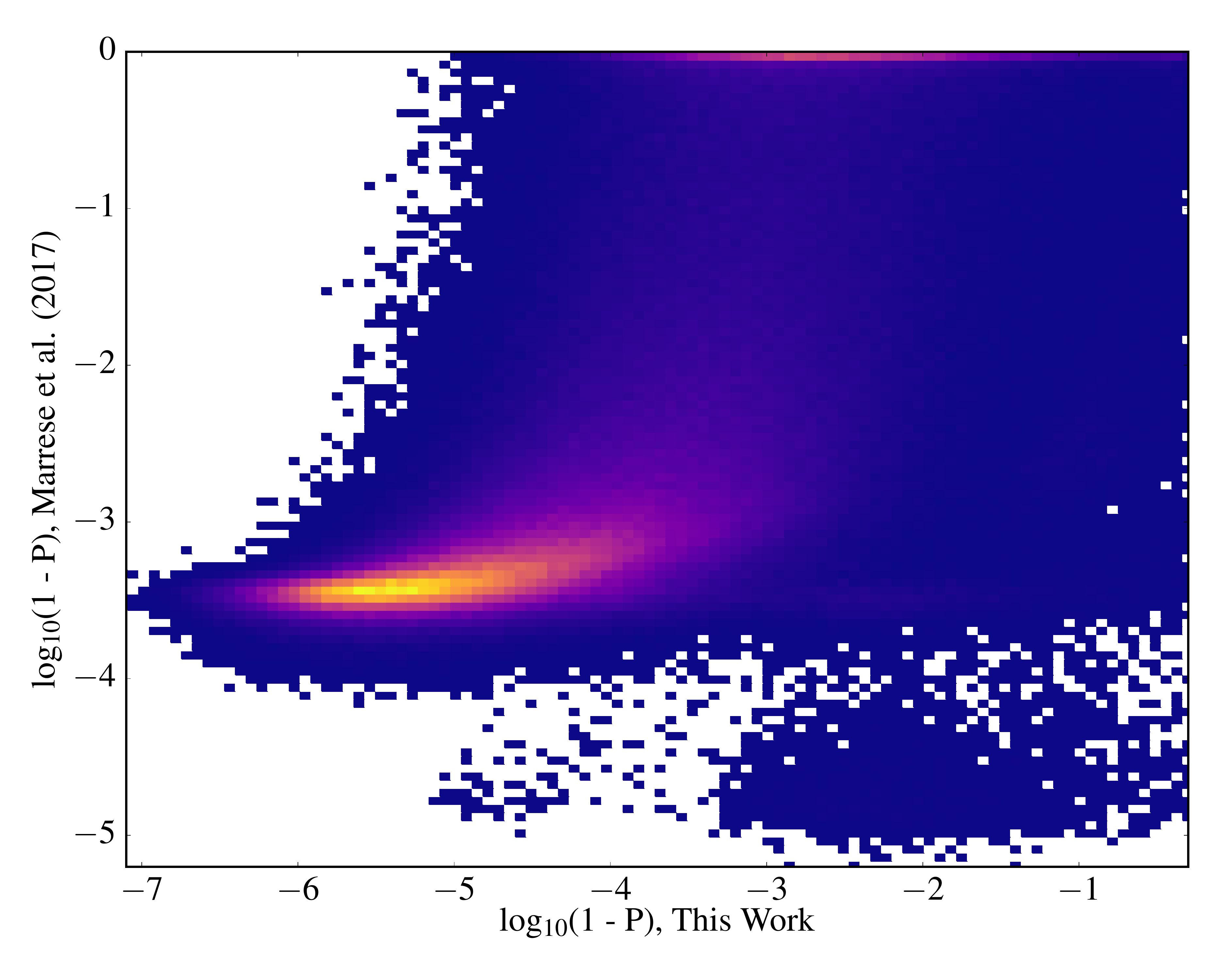}
    \caption{Comparison between match probabilities of \textit{Gaia} DR1 sources, as calculated using the method outlined in Section \ref{sec:empauf} and the probability calculated from the ``Figures of Merit'' as quoted by \citet{2017A&A...607A.105M}. The colour scale shows the number of stars in each two-dimensional bin, with blue representing low counts and yellow high counts. The astrometric broadening of the Gaussian uncertainties used by \citet{2017A&A...607A.105M} lead to a plateauing of the probabilities for the most certain matches, with as large as four orders of magnitude difference between the confidence with which the two methods reject the null hypothesis. The increase in probability for the small group of less certain matches is likely caused by the ``island'' creation process used by \citet{2018MNRAS.473.5570W}, leading to an increase in certainty of hypotheses when multiple pairs are simultaneously fit between the two catalogues.}
    \label{fig:gaiadr1comp}
\end{figure}

The effect in question can be explained as follows. The astrometric uncertainty broadening in turn reduces the maximum probability density of the Gaussian, being a normalised function, which has implications for null hypothesis testing. To test this we obtained the \textit{Neighbourhood} results for the \textit{Gaia}-\textit{WISE} matches from \citet{2017A&A...607A.105M}. We converted these scores to ``Figures of Merit'' ($FoM$), multiplying the figures of merit by a factor 3600 (P. Marrese, priv. comm.). In six cases there were multiple \textit{mates} for \textit{Gaia} sources, for which we picked the largest $FoM$ (see \citealp{2017A&A...607A.105M} for details).  We then obtained the ``reliability'' \citep{Sutherland:1992aa}, or the normalised probability of the pairing hypothesis, including the null hypothesis (or the two sources being uncorrelated and detections of differing objects) by

\begin{equation}
P(r) = (1 + (FoM(r))^{-1})^{-1}.
\label{eq:marreseprob}
\end{equation}
We then compared the probability obtained using the method laid out in Section \ref{sec:empauf} to those given as part of \textit{Gaia} DR1. 89\% of sources in each our cleaned \textit{Gaia} DR1 catalogue are shared with that provided by \citet{2017A&A...607A.105M} -- likely caused by differing quality cuts -- and their probabilities are compared in Figure \ref{fig:gaiadr1comp}. As can be seen, the broadening of the astrometric uncertainties leads to the most certain matches having a constant, but much lower, probability for the \citet{2017A&A...607A.105M} matches, compared with those using the method presented in this paper. This constant but reduced Gaussian probability density results in a reduction in the confidence with which we can reject the non-match hypothesis, by up to four orders of magnitude in some cases.

\subsubsection{Comparison With Forced Photometry}
\label{sec:compforcedphot}
The method we present here uses the catalogue cross-match, the assignment of individual sources from two catalogues, pre-merged within a given survey, as detections of one physical source. However, in recent years an alternative approach, typically referred to as ``forced photometry'', has been suggested for the creation of composite catalogues. Instead of creating two catalogues -- one per survey, with internal band merging -- all images across both catalogues have sources extracted simultaneously.

Extracting sources from all images at once allows for some level of crowding to be overcome, as the matching of an optical catalogue to a mid-IR catalogue allows for the extraction of sources resolved in the higher angular resolution optical dataset in the mid-IR data. The blended nature of a \textit{WISE} source can be forward-modelled from the detection of two SDSS sources, as in the case of \citet{Lang2016}. Indeed, the \textit{WISE} pipeline includes some limited source forward-modelling from its three shorter wavelength bandpasses to its $W4$ sources as its detection algorithm fits all four wavelengths simultaneously. It therefore allows for the possibility of the detection and characterisation of the fluxes of two sources resolved at the resolution of $W1$ but not at the resolution of $W4$. Detections -- in all four \textit{WISE} bandpasses -- are, in essence, regions of images with sufficient quadrature sum signal-to-noise ratios (see section 4.4b of \citealp{Cutri:2012aa}, building on the method described by \citealp{Szalay1999}, for more details).

This forward-modelling of sources detected in the best available individual catalogue passband can allow, provided the PSF is well understood and can be precisely modelled, for the recovery of additional sources in the lower resolution dataset. However, forced photometry trades this advantage off with other caveats. These drawbacks must be considered before a decision can be made as to which method is more appropriate for the given use case.

The first consideration is whether the two catalogues are seeing the same skies, or if they differ sufficiently in angular resolution, epoch, wavelength coverage, etc., to have very little overlap in the physical classes of sources they can detect. In the case we consider here, of one optical catalogue being merged with a dataset consisting of mid-IR detections, we may not detect the faint \textit{Gaia} sources causing the crowding in the \textit{WISE} images. We would therefore fail to forward-model their presence in a \textit{WISE} forced photometry catalogue. Forced photometry would, in these situations, throw away the knowledge that these sources, not detectable in a sparse sky, have perturbed the astrometry of the \textit{WISE} source and we would therefore lose the knowledge that the source is flux contaminated, unique to the method of modelling systematic perturbation due to crowding outlined in this paper.

dditionally, forced photometry is most advantageous when knowledge of the flux of sources detected in the highest angular resolution image from lower angular resolution images is required. Here, as detailed by \citet{Lang2016}, it may be possible to extract \textit{WISE} fluxes for sources detected at optical wavelengths. However, in the case we are considering in this paper, the question has been reversed: we wish to know the optical counterpart to detected \textit{WISE} sources. Forced photometry is therefore an asymmetric method, working from high to low angular resolution, and not applicable for the composite catalogue we are presenting here.

Finally, the consideration must be made as to the level of computational investment available. The forced photometry technique requires significant re-analysis of raw survey images, while the catalogue cross-match can be used on publicly available datasets, as in the case of \textit{Gaia} and \textit{WISE} here. The additional resources required to achieve detections using the alternative technique must be weighed carefully against the improvements offered.

The forced photometry method can provide improved detections in crowded regions under certain circumstances. However, these circumstances must be considered with care, as the method can also potentially produce worse results than the simpler survey-based catalogue creation and subsequent cross-match method. In the case we are interested in here, of the determination of the \textit{Gaia} counterparts to \textit{WISE} detections, the cross-match is the better method.

\subsubsection{Perturbation Offset Determination Comparison}
\label{sec:perturbcenter}
In this paper we deal with the effects of contaminant star perturbation by calculating flux-weighted centroid shifts to the central source, following \citet{2017MNRAS.468.2517W}. The applicability of these centroid shifts depends on the data reduction scheme applied to the images of the given observations. The flux-weighted centroid scheme is appropriate when positions have been found by centroiding, usually followed by aperture photometry to calculate the flux of the detection. However, there are data reduction schemes where PSF fitting is undertaken to calculate source fluxes and positions, the main method utilised to reconstruct sources in the \textit{WISE} data releases. In this instance the difference between two PSFs -- the bright source and the faint contaminating source -- and a slightly brighter, slightly shifted PSF representing the blended object should be minimised when evaluating potential perturbations. \citet{2018MNRAS.476.4372} use this method to explore the effects of confusion on the orbits of S-stars in the central few square arcseconds at the Galactic centre.

However, as \citet{2018MNRAS.476.4372} show, the analytical approximation to this minimisation only gives good agreement to the full solution for $\Delta m \geq 3$, or flux ratios less than approximately 6\%. Using the full numerical solution is computationally intractable for large-scale catalogue matching, and thus if this alternate method is considered the analytical expression would have to be used. For stars sufficiently bright that this inequality is valid, with typical contaminating sources at least three magnitudes fainter than the central source, we found that the centroiding and PSF fitting methods produce empirical AUFs that are in reasonable agreement with one another. We tested the offset perturbations produced by both methods against \textit{Gaia}-\textit{WISE} separation distributions for \textit{WISE} stars $131 \leq l \leq 138$, $-3 \leq b \leq 3$, $W1 = 10$, $\sigma_\alpha = 0.038"$, and $N = 0.263\,\mathrm{mag}^{-1}\,\mathrm{deg}^{-2}$. Both methods produced AUFs which fit the non-Gaussian tails to the separations, with fits to the full cross-match separations of $\chi_\nu^2 \simeq 1.5$ for both the PSF fitting method and the flux-weighted centroid method, with zero free parameters.

\textit{WISE} suffers extreme levels of crowding, however, and is potentially flux contaminated on the order of 15\% for stars as bright as $W1\simeq12$ (see Section \ref{sec:spitzerresolve}). Thus in regions of extreme crowding, or catalogues that are especially affected by crowding, such as \textit{WISE}, the flux-weighted centroid method produces AUFs much closer to the distribution of source separations than the analytical expression to the PSF fitting method derived by \citet{2018MNRAS.476.4372}. We tested this using the same sky region and normalising density as before, but with \textit{WISE} stars $W1 = 15$ and $\sigma_\alpha = 0.093"$. We found the flux-weighted centroid offset calculations produce offset distributions that result in an AUF with $\chi_\nu^2 \simeq 1.6$ when compared with the cross-match separation distribution, as previously, but the analytical approximation to the PSF fitting resulted in a much larger goodness-of-fit, $\chi_\nu^2 \simeq 6.5$. Therefore, while the flux-weighted centroiding method does not reflect the data reduction process as closely as the PSF fitting method, it produces AUFs in good agreement to the separations seen across all magnitudes. The analytical approximation to PSF fitting description does not hold for the majority of the \textit{WISE} stars, however. We therefore use the flux-weighted centroid method for the creation of the \textit{Gaia}-\textit{WISE} matches we present here.

\subsection{Photometry Differences}
\label{sec:photdiffs}
So far we have discussed the effects faint, hidden stars have on the astrometric positions of sources. Simultaneously, they also introduce additional flux to the central source. In this section we will discuss the effect crowding has on the photometry of blended sources, showing that the correct treatment of the astrometry of sources can reveal the introduction of additional brightness into these perturbed sources.

\subsubsection{The Effect of Perturbation on \textit{WISE} Brightnesses}
\label{sec:aufcomp}
The first test we can do to examine the effect crowding has on the flux contamination is to compare our two matching cases. We have, in effect, two distributions in our dataset of counterparts returned by the empirical AUF matching process used in Section \ref{sec:empauf}. First, we have those objects whose astrometric separations in \textit{Gaia} and \textit{WISE} are compatible with a Gaussian AUF, and would therefore have been matched in the Gaussian-based match in Section \ref{sec:gaussauf}. Second, we have the subset of objects with AUFs incompatible with a Gaussian, perturbed to the level that they were rejected by the Gaussian AUF matching process. These objects must have a hidden contaminant of sufficient brightness offset from the central source by a large enough radius such that the flux-weighted average position is beyond that allowed by the Gaussian AUF. In this subsection we will contrast these two subsets, to examine the effect this high level of perturbation has on the measured \textit{WISE} fluxes.

\begin{figure}
    \centering
    \includegraphics[width=\columnwidth]{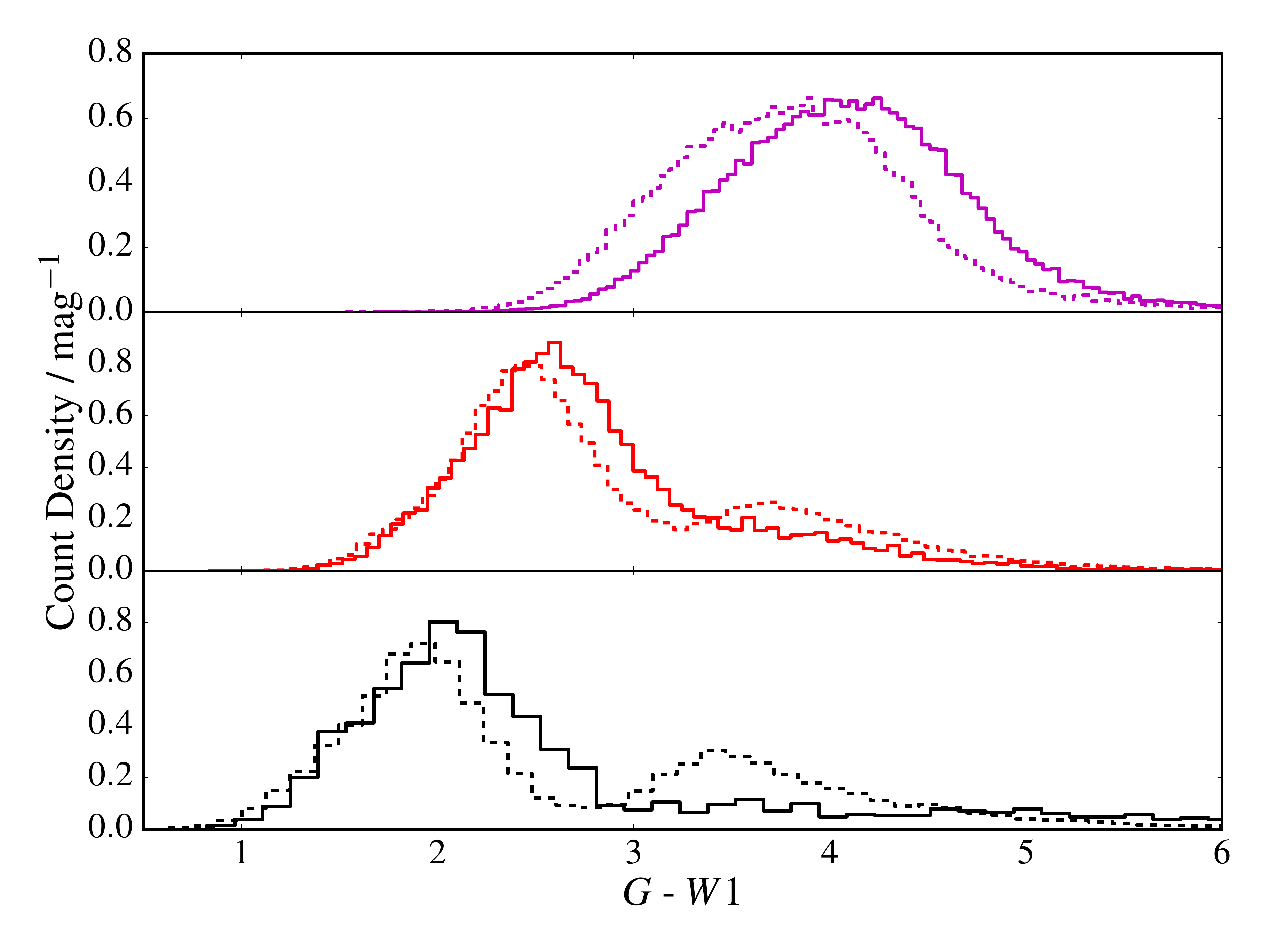}
    \caption{The $G-W1$ colour of \textit{Gaia}-\textit{WISE} matches that were paired using Gaussian-based AUFs (dashed lines) and the additional \textit{Gaia}-\textit{WISE} matches recovered using an empirical AUF (solid lines). Shown are the matches for stars with $12 \leq G \leq 14$ in black (bottom panel), $15 \leq G \leq 16$ in red (middle panel), and $19 \leq G \leq 20$ in magenta (top panel). The shift in $G-W1$ colour for those additional, empirical-only matches increases with increasing $G$ magnitude, suggesting an increasing $W1$ contamination. The average $W1$ magnitude is 0.27 magnitudes brighter for the non-Gaussian matches compared to those that are matched with a Gaussian AUF, implying $\simeq 28\%$ flux contamination, comparable to the average contamination seen in the constructed empirical AUFs.}
    \label{fig:gaussempcomp}
\end{figure}

Comparing the two distributions the immediate difference is the number of returned matches. The Gaussian-based match returns 58\% of the matches returned by the empirical AUF. However, more significant is the statistical relationship between the $G$ magnitude and \textit{WISE} bands (e.g., $W1$). Figure \ref{fig:gaussempcomp} shows the distribution of $G-W1$ colours for those objects recovered with a Gaussian-based match (dashed lines), and the additional objects that are paired when the empirical AUF is employed (solid lines), for several slices in \textit{Gaia} magnitude. With increasing $G$ magnitude, the $G-W1$ colour shift between the matches obtained with the purely Gaussian AUF and the additional empirical AUF-only matches increases. The objects gained when using an AUF that includes large, non-Gaussian wings are on average 0.27 magnitudes brighter in $W1$ for the same \textit{Gaia} magnitude than those recovered with a Gaussian AUF. This implies that the average flux contamination leading to these large wings, in those objects not captured by the Gaussian AUF due to sufficient flux contamination, is approximately 27\%, similar to the average flux contamination of 22\% seen in the empirical AUFs created in Section \ref{sec:empauf}.

To test further whether there was a correlation between photometric contamination and astrometric perturbation we divided the set of additional matches into two subsets, split by median sky separation. Fitting the $G-W1$ relationship of both halves of the gained matches we found a trend with $W1$ magnitude. At faint magnitudes ($W1\simeq16$) there is an inverse trend with match separation, with objects at smaller match separations exhibiting systematically more flux contamination. However, at increasingly bright magnitudes ($W1\lesssim13$) the objects with high astrometric perturbation are on average more flux contaminated than those objects that do not show high perturbation. At faint magnitudes, and thus high stellar densities, there are multiple contaminants in each \textit{WISE} PSF (\citealp{2017MNRAS.468.2517W}). As additional flux contamination from increasing numbers of contaminant stars is added, the overall flux-weighted centroid will tend towards zero. Therefore, in this high effective density regime the highest perturbations are seen in sources with lower levels of flux contamination, caused by a smaller number of faint sources. However, for brighter objects the effective stellar density is reduced, which leads to on average one contaminant that can affect the recorded position. This then results in a situation where there is a correlation between measured offset and contaminant brightness, as observed.

\subsubsection{Resolving Contaminants with \textit{Spitzer}}
\label{sec:spitzerresolve}
To confirm whether our sources are contaminated, we can examine the matches in a higher angular resolution dataset. \textit{WISE}'s $W1$ and $W2$ bands have very similar coverage to \textit{Spitzer}'s IRAC \citep{2004ApJS..154...10F} 3.6$\mu$m and 4.5$\mu$m bands, offering a resolution of $\simeq$2" FWHM. We therefore obtained \textit{Spitzer} Galactic Legacy Infrared Mid-Plane Survey Extraordinaire (GLIMPSE) data in the region $131 \leq l \leq 138$, $0 \leq b \leq 2$, and constructed empirical AUFs (see Section \ref{sec:empaufoveralldependence}) for the two IRAC filters available. We then performed a probability-based cross-match to \textit{Gaia} DR2 sources in the same region, as detailed by \citet{2018MNRAS.473.5570W}. We assumed that stars in both mid-infrared datasets that matched to the same \textit{Gaia} object were the same source detected at two different epochs in the two catalogues. We selected stars $11.5 \leq W1 \leq 12$, bright enough that \textit{WISE} sources are not entirely dominated by contamination, allowing for comparison between contaminated and uncontaminated sources.

\begin{figure}
    \centering
    \includegraphics[width=\columnwidth]{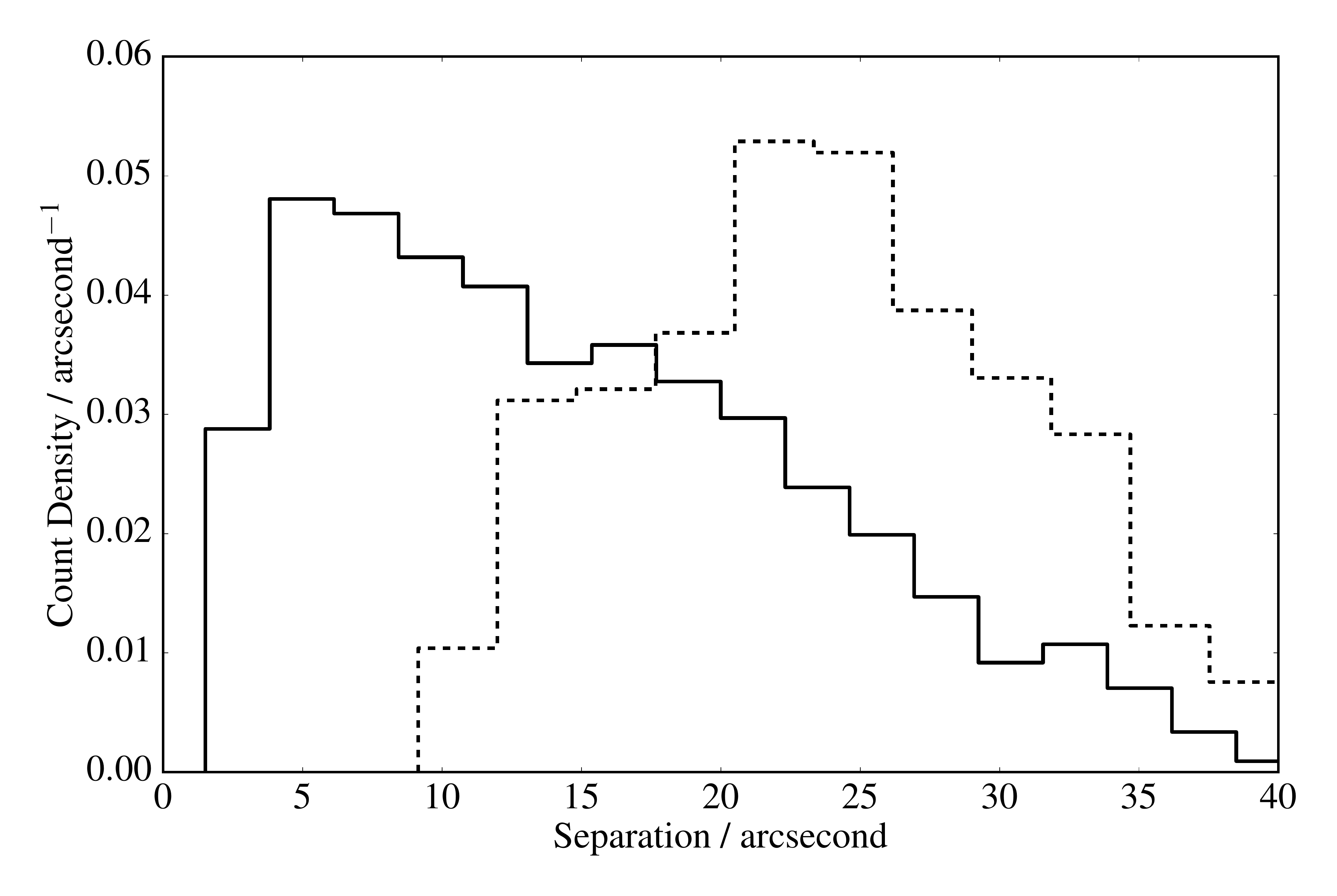}
    \caption{The intra-catalogue nearest neighbour distances for two samples of \textit{Spitzer} stars. Shown are those stars with a common \textit{Gaia} source to \textit{WISE} sources $11.5 \leq W1 \leq 12$. The two cases are those matches where the \textit{WISE} matches are unlikely to be contaminated ($P_\mathrm{contam} \leq 0.25$; dashed lines), and the case where the \textit{WISE} objects have a high probability of contamination ($P_\mathrm{contam} \geq 0.85$; solid lines). The faintest magnitude for intra-catalogue separation consideration was limited to 15, to account for \textit{Spitzer}'s fainter completeness limit. The \textit{Spitzer} detections of uncontaminated \textit{WISE} objects share a similar nearest neighbour distance distribution with both contaminated and uncontaminated \textit{WISE} sources. However, the \textit{Spitzer} nearest neighbour distribution for contaminated \textit{WISE} objects shows a much smaller average offset, with \textit{Spitzer} resolving the \textit{WISE} contaminants.}
    \label{fig:wisespitznndist}
\end{figure}

We obtained two subsets of these common \textit{Gaia} matches: likely uncontaminated \textit{WISE} objects and likely contaminated \textit{WISE} objects, based solely on our \textit{Gaia}-\textit{WISE} matching. These correspond to $P_\mathrm{contam} \leq 0.25$ and $P_\mathrm{contam} \geq 0.85$ (equation \ref{eq:contamprob}) respectively. Once we had obtained these four subsets (\textit{WISE} and \textit{Spitzer} objects which correspond to both contaminated and uncontaminated \textit{WISE} objects), we found the intra-catalogue separation (i.e., the distance to the nearest \textit{WISE} object for a given subset of \textit{WISE} objects). We limited the intra-catalogue search to stars with brightnesses $m \leq 15$ in both catalogues, allowing for consistent testing. Without the magnitude limit \textit{Spitzer}'s fainter completeness limit would otherwise have resulted in a smaller average offset than for that of \textit{WISE}, caused by an increase in the number of stars in any given region. The distribution of intra-catalogue separations for \textit{Spitzer} is shown in Figure \ref{fig:wisespitznndist}.

For the \textit{Spitzer} objects corresponding to uncontaminated \textit{WISE} objects (dashed line), the distribution of separations to the nearest intra-catalogue object (i.e., the nearest other \textit{Spitzer} detection) corresponds to the typical distance between sources at the given stellar density, approximately 25". This gives good agreement with both the contaminated and uncontaminated \textit{WISE} intra-catalogue distributions. However, the \textit{Spitzer} objects which correspond to contaminated \textit{WISE} objects (solid line) show a different distribution. With the better angular resolution \textit{Spitzer} has the ability to resolve two objects previously blended in \textit{WISE}. The nearest \textit{Spitzer} neighbour is therefore likely to be the hidden \textit{WISE} contaminant, as shown by a distribution skewed towards separations $\lesssim 10"$.

This resolving of contaminants is further confirmed when the magnitude differences between the \textit{WISE} and \textit{Spitzer} objects are compared for the two sources, similar to Section \ref{sec:aufcomp}. For the uncontaminated \textit{WISE} objects, the median $W1 - [3.6]$ colour is 0.005 magnitudes, while the subset of sources with significant \textit{WISE} contamination have a median $W1 - [3.6]$ of -0.141 magnitudes. This implies that, even as bright as $W1 \simeq 12$, some \textit{WISE} sources are suffering flux contamination on the order of 15\%.

\subsection{The Effects of Invisible Perturbants}
\label{sec:invispertub}
While there is good agreement between the empirical AUF constructed following the method laid out in Sections \ref{sec:constructempauf} and \ref{sec:empaufoveralldependence} and the distribution of separations between sources in the two catalogues, we can highlight here the effect of not including a more detailed treatment. The red dashed line in Figure \ref{fig:W1fitzmbreak} shows the empirical AUF obtained if the full treatment of the differential source counts is not taken into account (i.e., $N z^m$ is assumed to continue to arbitrarily faint magnitudes). As can be seen, this AUF does not fit the distribution of separations correctly; however, the magnitude of the central sources is almost at the sensitivity limit of the survey. 

This means that the vast majority of sources affecting the AUF and the perturbation of the central sources would not be detected by the survey in a sparse field. This highlights the importance of the correct treatment of the density of faint contaminants. If treated correctly, the effects of otherwise ``invisible'' stars can be seen indirectly in their influence on brighter objects.

\subsection{Circular Symmetry in Empirical AUF Creation}
\label{sec:circsymmetryauf}
The formalism given here for the creation of empirical AUFs implicitly assumes circular symmetry. We have assumed a circular PSF in the previous sections, and for the discussion in Sections \ref{sec:empaufeffect} and \ref{sec:photdiffs} we additionally assume the astrometric uncertainties are circular (i.e., $\rho = 0$, $\sigma_\alpha = \sigma_\delta$). This assumption holds for the majority of sources, as ground-based surveys should have circular PSFs and thus circular centroiding uncertainties. Space-based observations, such as those for \textit{WISE}, can have off-axis correlations in their PSFs, and thus position uncertainties, however. In practice, this effect is limited and 90\% of the \textit{WISE} data discussed here have orthogonal sky axis uncertainties that deviate from circular by less than 10\%. Therefore, while the convolution of the distribution of perturbations and a Gaussian preserving the full covariance matrix is possible, the loss of information is negligible, vastly outweighed by the simplifications the assumption allows.

\begin{figure}
\centering
	\includegraphics[width=\columnwidth]{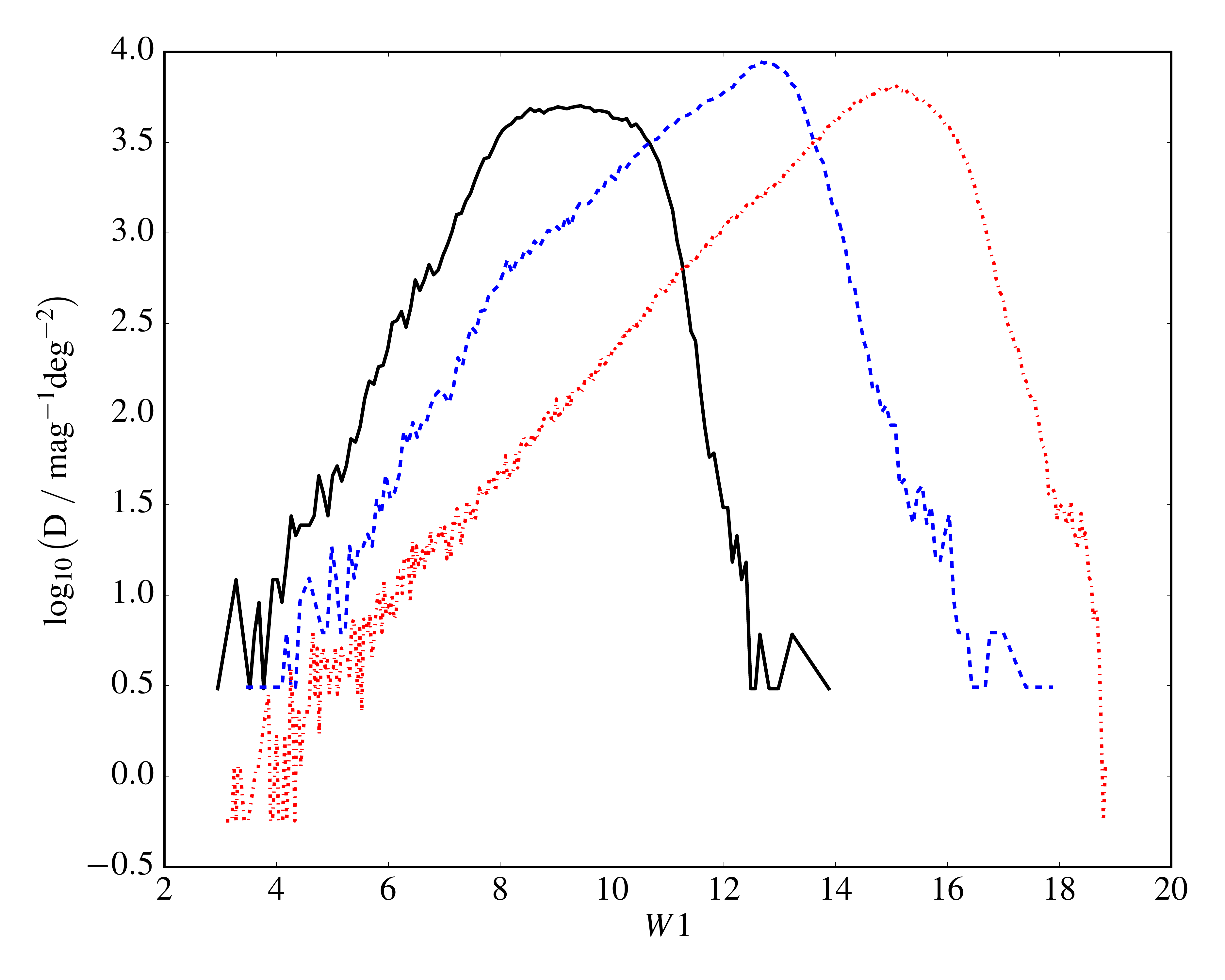}
	\caption{\textit{WISE} differential source counts. The red dotted line is the differential source counts for the region of the Galactic plane around $l=135$, $b=0$ discussed in Section \ref{sec:empaufeffect}. The blue dashed and black solid lines are the differential source counts of 4 square degree regions of the Galactic plane at $355 \leq l \leq 357$, $5 \leq b \leq 7$ and $0 \leq l \leq 2$, $0 \leq b \leq 2$ respectively. The blue dashed and red dotted lines show relationships that follow a $z\simeq2$ scaling law. However, the crowding in the inner region of the Galactic centre is so extreme that the brightest sources are flux contaminated to such an extent that they no longer follow a geometric scaling law.}
	\label{fig:wiselvariation}
\end{figure}

\subsection{Extreme Crowding}
\label{sec:extremecrowding}
The method we have outlined in this paper accounts for the blending of sources, including the effects the brightening of the brightest source has on its astrometry. However, one of the assumptions made was that the local density of each source could be calculated from a consistent geometric scaling relationship. As shown in Figure \ref{fig:wiselvariation} (cf. Figure \ref{fig:wiseview0}), this assumption may not necessarily hold in regions of extreme crowding. The black solid line shows \textit{WISE} differential source counts of a 4 square degree region of the inner Galactic centre ($l=0, b=0$). Compared with the blue dashed and red dotted lines, representing differential source counts at $l=355$, $b=5$ and $l=135$, $b=0$ respectively, the Galactic centre suffers such extreme flux contamination that its brightest sources no longer follow a geometric scaling relationship. We therefore chose to remove from the catalogue of pairings presented in Section \ref{sec:gplanematch} any sources in the region $\lvert l \lvert\ \leq 10$, $\lvert b \lvert\ \leq 5$. In cases where the density of sources, i.e., number of stars per PSF, is extreme, we recommend analysing the differential source counts for the catalogue in question to ensure the assumptions made about the scaling law relationship are still valid.

\subsection{Extensions to the AUF}
\label{sec:aufextensions}

\subsubsection{Extending the Empirical AUF to Additional Systematic Perturbations}
\label{sec:extendaufperturb}
In this work we have chosen to only include the systematic effects of crowding in our AUF treatment, being the most dominant source of non-Gaussianity in the \textit{WISE} AUFs \citep{2017MNRAS.468.2517W}. However, AUFs can include any source of systematic perturbation without loss of generality. We can include other effects such as proper motion, described by

\begin{equation}
h_\mathrm{tot} = h_\mathrm{pure} * h_\mathrm{offsets} * h_\mathrm{pm}.
\end{equation}
Here $h_\mathrm{pm}$ is a probability density function describing the statistical distribution of proper motions for the catalogue in question. This has the potential to model the effects of proper motion on a large scale, in cases where individual measurements are unavailable. For example, stars fainter than \textit{Gaia} in the next generation of photometric surveys, such as LSST, will likely lack robust individual proper motion measurements. Modelling their effects will therefore rely on such large scale statistical proper motion simulations.

The ability to include the distribution of the proper motions of sources, rather than merely inflating the astrometric uncertainty of the position centroiding (e.g., \citealp{2017A&A...607A.105M}), allows for a more realistic treatment of these systematic perturbations to source positions. It can be extended to be a function of multiple parameters of the catalogue -- the primary one being brightness, with fainter objects having smaller proper motions on average -- and does not erase the knowledge of the original positional precision. There are several cases in the literature where the motion of sources is included in the cross-matching of catalogues. \citet{Pineau:2017aa} include an appendix discussing extending their maximum-likelihood Gaussian AUF treatment to the inclusion of the motion of sources between catalogue epochs. Similarly \citet{2010ApJ...719...59K} extend the Gaussian AUF Bayes factor method of \citet{Budavari:2008aa} to account for unknown proper motions, including a more detailed treatment of the likely astrophysical proper motions of the sources as a prior term. We believe the inclusion of the proper motion offset term as part of the likelihood, inside the combined AUF, to be a more intuitive interpretation to the positional offset between catalogue source detections. It simply continues the extension to non-Gaussian perturbations, adding all terms required to correctly interpret the separations between counterpart detections to astrophysical sources.

If motions for individual sources are known, perhaps due to indivdually known proper motions or an absolute catalogue position offset relative to the second catalogue, then $h$ could simply be a delta function. This would result in the convolution being evaluated with a simple shift in astrometric coordinates, as $(f * \delta)(t) = f(t)$. In practice, however, this is most likely simpler to handle before beginning the cross-match, during the creation of a given catalogue. Indeed, we applied a proper motion shift to the \textit{Gaia} DR2 dataset used in this paper to account for the five year epoch difference between its observations and the typical epoch of the \textit{WISE} data. We linearly extrapolated the proper motions of sources in each orthogonal axis, as given by

\begin{align}
\begin{split}
\alpha_\mathrm{new} &= \alpha - 5 \mathrm{year} \cdot \mu_\alpha\left[\cos(\delta)\right]^{-1}\\
\delta_\mathrm{new} &= \delta - 5 \mathrm{year} \cdot \mu_\delta,
\label{eq:pmdrift}
\end{split}
\end{align} 
as \textit{Gaia} DR2 is recorded in epoch 2015.5 and the \textit{WISE} mission operated during 2010.

\begin{figure}
\centering
	\includegraphics[width=\columnwidth]{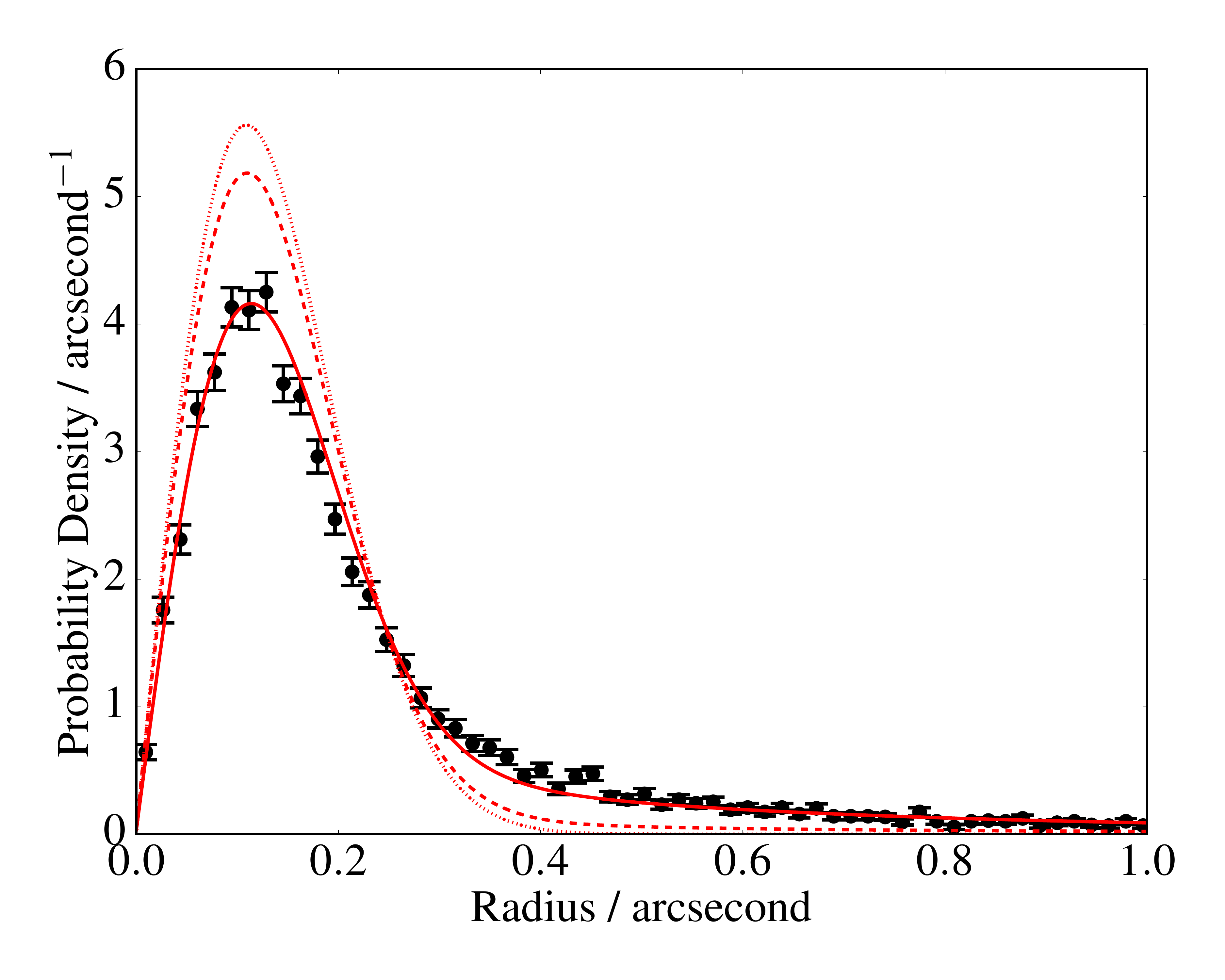}
	\caption{\textit{Gaia}-\textit{WISE} matches for the Galactic North Pole, $b \geq 75$, for $N = 0.009-0.019\, \mathrm{mag}^{-1} \mathrm{deg}^{-2}$, $W1 = 15.47-15.53$, and $\sigma_\alpha=0.08-0.14"$, shown in the black errorbars. The Rayleigh distribution (the representation of a two-dimensional Gaussian in one-dimensional radial coordinates) of the given astrometric uncertainty is shown as a red dotted line. The empirical AUFs including the effects of perturbation from \textit{WISE} Galactic star and galaxy counts are shown as the red dashed and solid lines, respectively. The low density of Galactic sources leads to little perturbation, but the order-of-magnitude higher galaxy counts leads to an AUF in agreement with the distribution of separations.}
	\label{fig:wisepolesep}
\end{figure}

\subsubsection{Extensions to Extra-galactic Source Contamination}
\label{sec:extragalacticcontamination}
In this paper we have focussed on discussion of the effects of contamination on Galactic sources, focussing on sources with $\lvert b \lvert\ \leq 10$. These stars suffer much higher average crowding than those sources out of the plane of the Galaxy, and much more crucially need these effects taking into account. However, for catalogues at longer wavelengths with deep completeness limits, such as \textit{WISE}, faint galaxy source counts will play a role in the perturbation of brighter detections. These perturbations are entirely analogous to the Galactic contamination dealt with in Section \ref{sec:empauf}, and extra-galactic sources contribute to the perturbation of sources in the Galactic plane. However, the stellar densities at these Galactic latitudes are much higher than the typical galaxy counts. Additionally, the significant levels of interstellar extinction most significantly affect extra-galactic sources, decreasing their brightnesses more than those of the stars in the Galaxy, further exacerbating the differential source count discrepancy. The contribution of extra-galactic sources to the perturbation of the \textit{WISE} sources considered in this paper is therefore small. More generally, however, these additional sources from outside the Galaxy can significantly affect the AUFs of these faint, long wavelength catalogues.

This effect is highlighted in Figure \ref{fig:wisepolesep}, where the distribution of nearest neighbour matches for Galactic North Pole \textit{Gaia}-\textit{WISE} stars, $b \geq 75$, is shown in black errorbars with $N=0.014\, \mathrm{mag}^{-1} \mathrm{deg}^{-2}$ (calculated using the differential star counts via equation \ref{eq:solveforn}), $W1 = 15.5$, and $\sigma_\alpha = 0.11"$. For reference a pure Gaussian AUF of the quoted astrometric uncertainty is plotted as the red dotted line. The empirical AUF calculated when taking into account the effects of Galactic \textit{WISE} stars, using the TRILEGAL differential source counts (see Sections \ref{sec:constructempauf} and \ref{sec:empaufoveralldependence}), is shown as a red dashed line. As can be seen, the Galactic stellar density at the Galactic pole is low ($N$ being some factor of 25 smaller than that typical of the Galactic plane), leading to low astrometric perturbation.

At faint mid-infrared magnitudes, however, the density of galaxies can reach a factor of 10 higher than that of Galactic sources (e.g., figure 7 of \citealp{2017ApJ...836..182J}). Constructing the differential \textit{WISE} galaxy count using the galaxy counts of \citet{2017ApJ...836..182J} (see Section \ref{sec:empaufzmdependence} for discussion on construction multiple geometric scaling law relationships), our galaxy contaminant empirical AUF is shown in Figure \ref{fig:wisepolesep} as the red solid line. These perturbations produce an AUF in agreement with the distribution of match separations. Therefore, when considering faint, long wavelength detections it is critical that the effects of both Galactic and extra-galactic sources are considered.

\section{Conclusions}
\label{sec:conclusions}
We presented an analysis of the effects of unresolved contaminant stars on the cross-matching of the \textit{Gaia} and \textit{WISE} photometric catalogues. We detailed a treatment of the astrometric uncertainty functions which is capable of folding in these systematic astrometric perturbations in Section \ref{sec:empauf}. Comparisons between the ensemble of pairings produced by a probability-based matching process using Gaussian AUFs and the new empirical AUFs were carried out. It was found that without the inclusion of the effects of contamination one in every two \textit{Gaia}-\textit{WISE} match is rejected. We also detailed the results of a number of test matches, analysing the match rates, false match rates, and effects on the probabilities obtained in Section \ref{sec:gplanematch}.

In addition to discussing the effects these unresolved objects have on the astrometry, in Section \ref{sec:discussion} we considered the effect crowding has on the measured photometric magnitudes. We found that \textit{WISE} objects perturbed sufficiently to be entirely incompatible with a Gaussian AUF are on average 30\% brighter than those objects with small astrometric perturbations. Additionally, we compared the \textit{WISE} matches to \textit{Spitzer} detections, using the superior angular resolution of \textit{Spitzer} to resolve the \textit{WISE} contaminants. The ability to resolve the previously blended \textit{WISE} contaminants leads to a skewed intra-\textit{Spitzer} separation distribution.

We also provided a catalogue of the probability-based matches between \textit{Gaia} DR2 and \textit{WISE} for the Galactic plane $\lvert b \rvert \leq 10$ in Section \ref{sec:gplanematch}. This catalogue provides the probability of both the \textit{Gaia}-\textit{WISE} pairing and the probability that the \textit{WISE} object suffer contamination above the 1\% flux level. Modelling the effects of hidden contaminants is important for correctly matching two detections which otherwise would have been assumed to be two unphysical individual detections. Moreover, it also allows for the selection of only objects without significant flux from additional sources, critical for comparisons to theoretical models.

\section*{Acknowledgements}
\label{sec:acknowledge}
The authors thank the referee for their useful comments and feedback, which led to an improved manuscript. TJW acknowledges support from an STFC Studentship. TN is funded by a Leverhulme Trust Research Project Grant. This work has made use of the SciPy \citep{scipy}, NumPy \citep{numpy}, Matplotlib \citep{matplotlib}, and F2PY \citep{f2py} Python modules, and NASA's Astrophysics Data System.

This publication makes use of data products from the Wide-field Infrared Survey Explorer, which is a joint project of the University of California, Los Angeles, and the Jet Propulsion Laboratory/California Institute of Technology, funded by the National Aeronautics and Space Administration. 

This work has made use of data from the European Space Agency (ESA) mission {\it Gaia} (\url{https://www.cosmos.esa.int/gaia}), processed by the {\it Gaia} Data Processing and Analysis Consortium (DPAC, \url{https://www.cosmos.esa.int/web/gaia/dpac/consortium}). Funding for the DPAC has been provided by national institutions, in particular the institutions participating in the {\it Gaia} Multilateral Agreement.

This work is based [in part] on observations made with the Spitzer Space Telescope, which is operated by the Jet Propulsion Laboratory, California Institute of Technology under a contract with NASA.




\bibliographystyle{mnras}
\bibliography{../../PostgradPapers_jr2.bib} 
\bsp	
\label{lastpage}
\end{document}